\documentclass[times,twocolumn,final]{elsarticle} 

\usepackage{makecell}
\usepackage{fancyhdr}  

\makeatletter
\def\ps@pprintTitle{%
 \let\@oddhead\@empty
 \let\@evenhead\@empty
 \renewcommand{\footrulewidth}{0pt} 
 \def\@oddfoot{\mbox{\textit{Preprint accepted for publication in Medical Image Analysis}} \hspace{1cm}} 
 \let\@evenfoot\@oddfoot}
\makeatother

\usepackage{framed,multirow}
\usepackage{amssymb}
\usepackage{amsmath}
\usepackage{latexsym}
\usepackage{adjustbox}
\usepackage{url}
\usepackage{xcolor}
\usepackage{hyperref}

\definecolor{newcolor}{rgb}{.8,.349,.1}

\usepackage{amssymb}
\usepackage{amsmath}
\usepackage{latexsym}
\usepackage{adjustbox}

\newcommand{\norm}[1]{\left\lVert#1\right\rVert}
\usepackage{url}
\usepackage{xcolor}

\usepackage{hyperref}

\newcommand{\argmin}{\mathop{\mathrm{arg\,min}}\nolimits}
\begin{document}


\begin{frontmatter}

\title{MBSS-T1: Model-Based Subject-Specific Self-Supervised Motion Correction for Robust Cardiac T1 Mapping}


\author[1]{Eyal Hanania\corref{cor1}}
\cortext[cor1]{Corresponding author}
\ead{EyalHan@campus.technion.ac.il}
\author[2]{Adi Zehavi-Lenz}
\author[4]{Ilya Volovik}
\author[2,3]{Daphna Link-Sourani}
\author[1]{Israel Cohen}
\author[2,3]{Moti Freiman}

\address[1]{Faculty of Electrical \& Computer Engineering, Technion - IIT, Haifa, Israel}
\address[2]{Faculty of Biomedical Engineering, Technion - IIT, Haifa, Israel}
\address[3]{The May-Blum-Dahl MRI Research Center, Faculty of Biomedical Engineering, Technion - IIT, Haifa, Israel}
\address[4]{Bnai Zion Medical Center, Haifa, Israel}


\begin{abstract}
Cardiac T1 mapping is a valuable quantitative MRI technique for diagnosing diffuse myocardial diseases. Traditional methods, relying on breath-hold sequences and cardiac triggering based on an ECG signal, face challenges with patient compliance, limiting their effectiveness. Image registration can enable motion-robust cardiac T1 mapping, but inherent intensity differences between time points pose a challenge. We present MBSS-T1, a subject-specific self-supervised model for motion correction in cardiac T1 mapping. Physical constraints, implemented through a loss function comparing synthesized and motion-corrected images, enforce signal decay behavior, while anatomical constraints, applied via a Dice loss, ensure realistic deformations. The unique combination of these constraints results in motion-robust cardiac T1 mapping along the longitudinal relaxation axis. In a 5-fold experiment on a public dataset of 210 patients (STONE sequence) and an internal dataset of 19 patients (MOLLI sequence), MBSS-T1 outperformed baseline deep-learning registration methods. It achieved superior model fitting quality ($R^2$: 0.975 vs. 0.941, 0.946 for STONE; 0.987 vs. 0.982, 0.965 for MOLLI free-breathing; 0.994 vs. 0.993, 0.991 for MOLLI breath-hold), anatomical alignment (Dice: 0.89 vs. 0.84, 0.88 for STONE; 0.963 vs. 0.919, 0.851 for MOLLI free-breathing; 0.954 vs. 0.924, 0.871 for MOLLI breath-hold), and visual quality (4.33 vs. 3.38, 3.66 for STONE; 4.1 vs. 3.5, 3.28 for MOLLI free-breathing; 3.79 vs. 3.15, 2.84 for MOLLI breath-hold). MBSS-T1 enables motion-robust T1 mapping for broader patient populations, overcoming challenges such as suboptimal compliance, and facilitates free-breathing cardiac T1 mapping without requiring large annotated datasets. \textit{Our code is available at \url{https://github.com/TechnionComputationalMRILab/MBSS-T1}}.

\end{abstract}

\begin{keyword}
T1 Mapping\sep Motion Correction\sep Myocardial Imaging\sep Deep Learning\sep Model-based Deep Learning\sep Free-breathing MRI
 \end{keyword}

\end{frontmatter}


\section{Introduction}
Quantitative cardiac T1 mapping is an advanced MRI technique designed to precisely measure intrinsic longitudinal relaxation time in myocardial tissue \cite{taylor2016t1}. This method is increasingly recognized as essential for assessing diffuse myocardial diseases such as inflammation, fibrosis, hypertrophy, and infiltration \cite{taylor2016t1,van1998quantitative}. ``Native'' cardiac T1 mapping, performed without a paramagnetic contrast agent, effectively detects myocardial conditions, including edema, iron overload, infarcts, and scarring \cite{schelbert2016state}. Furthermore, by allowing for the objective quantification of tissue characteristics, cardiac T1 mapping enables longitudinal tracking of changes in the myocardium. This feature is precious for clinical trials, where monitoring the efficacy of interventions over time is critical \cite{schelbert2016state}.

\begin{figure*}[t!]
\includegraphics[width=0.75\textwidth]{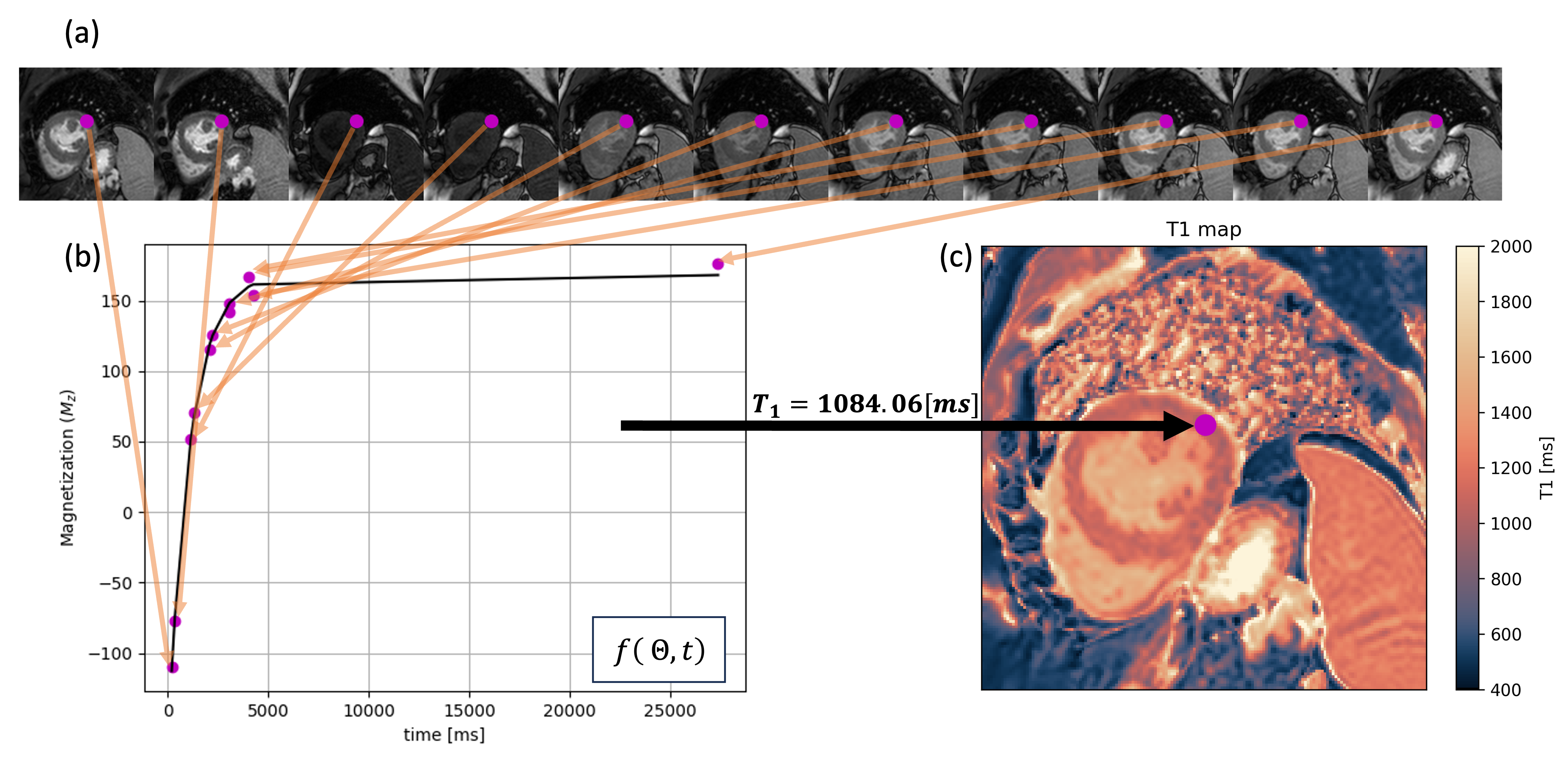}
\centering
\caption{Schematic representation of cardiac T1 mapping for a single voxel. 
(a) T1-weighted myocardial images acquired at \(N-1\) sequential time points. 
(b) Fitting an inversion recovery curve \( f\left(\Theta,t\right) \) of the longitudinal magnetization across time points \( t \) to estimate the corresponding parameters \( \Theta \). In this STONE sequence with the two-parameter model, \( \Theta = \{M_0, T1\} \) and \( f\left(\Theta,t\right) = M_0 \cdot \left(1 - 2 \cdot e^{-t/T1} \right) \), where \( M_0 \) represents the equilibrium magnetization, which is the tissue magnetization before any preparation, and \( T1 \) reflects the actual T1 value. 
(c) Computed T1 map visualizing the estimated T1 values across the myocardium.}

\label{fig:relaxation}
\end{figure*}

Figure~\ref{fig:relaxation} presents the process of generating accurate T1 maps. It begins with capturing a series of images at different time points. Each image captures the myocardium at a distinct point during the longitudinal relaxation. These images are then analyzed to calculate T1 relaxation times through signal model fitting, resulting in detailed T1 maps that reflect the myocardial tissue properties. However, heart motion, respiration, and spontaneous patient movements can cause significant distortions in T1 maps, affecting their reliability and clinical usefulness and potentially leading to incorrect diagnoses \cite{tilborghs2019robust}. Cardiac triggering based on an ECG signal is a well-established method for accounting for cardiac motion. Breath-hold sequences, such as the Modified Look-Locker Inversion Recovery (MOLLI) sequence and its variants \cite{roujol2014accuracy}, are often used to minimize respiratory motion artifacts. However, requiring patients to hold their breath presents practical challenges, as not all patients can fully adhere to these instructions. This breath-hold acquisition further limits the number of images that can be captured and makes cardiac T1 mapping unsuitable for patients who cannot tolerate breath-holding \cite{roujol2014accuracy}. 


Reducing motion artifacts and enabling free-breathing cardiac T1 mapping can be accomplished through image registration to align single-shot images taken at different time points. However, achieving precise registration is challenging due to inherent complexities in the image data. These complexities include contrast inversion, partial volume effects, and signal nulling for images captured near the zero crossing of the T1 relaxation curve (Figure~\ref{fig:relaxation}).

Despite advancements in classical and deep learning-based methods for motion correction and cardiac T1 mapping, significant limitations persist. Classical model-based approaches, while effective, are often slow due to their iterative nature \cite{tilborghs2019robust, xue2012motion}. On the other hand, deep-learning-based methods rely heavily on extensively annotated datasets, which are not readily available for all scanners and protocols \cite{gonzales2021moconet, yang2022disq}. Furthermore, these methods typically use either pairwise loss functions based on metrics such as mutual information to guide the registration process \cite{arava2021deep} or groupwise approaches such as the principal component analysis (PCA), which ignore the signal decay relaxation model \cite{zhang2024pca}. This omission may result in physically unlikely deformations due to changes in image contrast. These limitations hinder the practical applicability and robustness of current methods.

This work presents MBSS-T1, a subject-specific self-supervised, physically, and anatomically constrained deep learning model for simultaneous motion correction and cardiac T1 mapping from either breath-hold or free-breathing acquisitions. Our method integrates physical and anatomical insights to correct motion in cardiac T1 mapping. The main innovation of our approach lies in embedding the signal model directly into the network architecture within a subject-specific self-supervised framework. Additionally, we leverage a segmentation network pre-trained on the STONE dataset \cite{el2018nonrigid,DHEUAV_2019} to generate segmentations that guide the registration process, encouraging anatomically accurate alignment of the deformed images to a reference frame. This approach enables the generation of motion-robust cardiac T1 maps across various imaging protocols, eliminating the need for protocol-specific data collection, manual annotations, and extensive training. 
This work extends our previous supervised PCMC-T1 model \cite{hanania2023pcmc}, which required a large annotated dataset and extensive training, thereby limiting its generalization capacity across different acquisition settings, such as sequence type and the number of frames acquired.

We have demonstrated the added value of MBSS-T1 in free-breathing cardiac T1 mapping using a 5-fold experimental setup. This was done on a publicly available free-breathing quantitative dataset of 210 patients \cite{el2018nonrigid}, acquired using the STONE sequence \cite{weingartner2015free}, and an in-house dataset of 19 patients acquired using the MOLLI sequence \cite{roujol2014accuracy}, for both free-breathing and breath-hold scans. Our approach was compared to baseline methods for deep-learning-based image registration \cite{dalca2019unsupervised, hoffmann2021synthmorph}, as well as the T1 maps produced on the scanner.
We used the fit quality (r-squared), the Dice coefficient, and the Hausdorff distance as quantitative metrics to assess registration accuracy and clinical usability. We further demonstrated the clinical impact through qualitative expert radiologist assessments.

\section{Related Work}
Previous works aimed to address the challenge of motion-robust cardiac T1 mapping can be divided into classical approaches and more recent deep-learning-based approaches. Classical approaches 
have primarily relied on model-based techniques and iterative algorithms to mitigate the effects of intensity variations between the different frames. More recently, deep learning-based methods have emerged, leveraging large datasets and neural network architectures to enhance motion correction and cardiac T1 mapping accuracy. In the following, we review these works.

\subsubsection*{Classical Approaches:} Zhang et al. addressed intensity variations in their pairwise non-rigid elastic registration framework for cardiac T1 mapping by employing the normalized gradient fields difference as their similarity metric \cite{zhang2018cardiac}. Similarly, Roujol et al. presented an adaptive registration method for varying contrast-weighted images. This approach uses a local non-rigid registration framework based on optical flow to simultaneously estimate motion fields and intensity variations, effectively managing significant contrast differences. Additionally, the method incorporates a regularization term to constrain the deformation field through automatic feature tracking \cite{roujol2015adaptive}.

Guyader et al. introduced a groupwise image registration technique for quantitative MRI that utilizes a dissimilarity metric derived from an approximated form of total correlation. This method addresses the challenge of aligning multiple images simultaneously without relying on a reference image, thereby reducing registration bias and improving the consistency of quantitative measurements \cite{guyader2016total}. Tao et al. introduced a robust motion correction method for myocardial T1 and extracellular volume mapping using a PCA-based groupwise image registration technique, which registers all MOLLI frames simultaneously with a total-correlation-based metric \cite{tao2018robust}. Similarly, Huizinga et al. developed a PCA-based groupwise image registration technique for quantitative MRI (qMRI), including cardiac T1 mapping, to correct misalignment without a reference image, thus avoiding registration bias. Their method registers all images to a mean space by minimizing a PCA-derived cost function, assuming that a low-dimensional signal model can represent intensity changes in qMRI without needing details of the specific acquisition model \cite{huizinga2016pca}.

Segmentation-based registration techniques address intensity differences between images using anatomical structures to guide the registration. El-Rewaidy et al. utilized a non-rigid active shape model-based framework, registering T1-weighted images and computing residual motion by matching manually annotated myocardium contours. This method improves T1 map quality by addressing motion-induced errors in myocardial segments \cite{el2018nonrigid}.

A key limitation of the methods above is their failure to account for the signal model during registration, resulting in potentially non-physical corrected T1 maps. Model-based approaches have been developed to address this, significantly enhancing cardiac T1 mapping accuracy. Xue et al. proposed a fast variational non-rigid registration framework using synthetic image estimation to correct motion in myocardial T1 mapping. Motion-free synthetic images, constructed from initial T1 estimates, guide the registration to avoid motion artifacts \cite{xue2012motion}. Van Heeswijk et al. introduced a model-based alignment of Look-Locker MRI sequences for myocardial scar quantification, using exponential curve fitting errors as the registration metric \cite{van2013model}. Tilborghs et al. applied a robust motion correction method for cardiac T1 and ECV mapping, integrating data-driven initialization with model-based registration within a non-rigid framework. This method iteratively estimates signal model parameters, generates synthetic images, and uses these for accurate registration \cite{tilborghs2019robust}. However, their iterative nature makes these model-based approaches slower than non-iterative methods.

\subsubsection*{Deep-Learning Approaches:} Deep-learning approaches have also been proposed for motion correction by image registration. These networks typically learn a function that directly outputs a deformation field given an image pair. For instance, Morales et al. \cite{morales2019implementation} implemented and validated a three-dimensional convolutional neural network for estimating cardiac motion from Cine MRI images. Kustner et al. \cite{kustner2021lapnet} proposed a deep learning approach for fast and accurate non-rigid registration directly from undersampled k-space data of cardiac T1-weighted spoiled gradient echo images. Qi et al. \cite{qi2021end} developed a supervised end-to-end deep neural network capable of simultaneously performing 3D non-rigid motion estimation and motion-compensated reconstruction from undersampled cardiac k-space data. Hering et al. \cite{hering2019enhancing} improved cardiac motion tracking by integrating global semantic information from segmentation labels with local distance metrics, thereby enhancing the alignment of surrounding structures. Teed et al. \cite{teed2020raft} introduced RAFT, a deep neural network employing recurrent units to estimate optical flow in videos; however, it was not specifically designed for medical imaging applications. Despite their advancements, all these methods assume relatively consistent signal appearances between the frames being registered, which limits their applicability to cardiac T1 mapping, where inherent signal variations exist across frames.

Several deep-learning approaches have been proposed to address the inherent signal variations in cardiac T1 mapping. Arava et al. \cite{arava2021deep} introduced a recursive framework that cascades pairwise registration systems for motion correction in T1-weighted images, minimizing mutual information between each pair of images. Hanania et al. \cite{hanania2023groupT1} employed multi-image registration models optimized using a mutual-information-based loss function. Gonzales et al. \cite{gonzales2021moconet, gonzales2022fast} developed a multi-scale U-Net architecture to generate distance vector fields, incorporating warping layers to progressively deform feature maps from coarse to fine. This strategy enables the alignment of images with varying intensities while reducing motion artifacts through iterative refinement. Li et al. \cite{li2022motion} applied a sparse coding technique to extract contrast components from the images, followed by a self-supervised deep neural network (DNN) for image registration. In a related work, Li et al. \cite{li2021deep} integrated spatial-temporal and physical constraints with low-rank and sparsity regularization into a deep neural network for efficient cardiac T1 mapping, denoising, and artifact reduction.

Qin et al. \cite{qin2018joint} proposed a joint learning approach for motion estimation and segmentation, leveraging unsupervised features learned from large amounts of unannotated data to improve registration accuracy. Pan et al. \cite{pan2024virtual} and Hanania et al. \cite{hananiastylereg} adopted generative adversarial networks (GANs) to perform style transfer, generating virtual images at different inversion times from a single input image. By standardizing contrast across images, the GAN-based methods reduce sensitivity to intensity variations and enhance the alignment of images with varying contrasts.

Yang et al.~\cite{yang2022disq} introduced a sequential method to address contrast differences by initially disentangling intensity changes at different inversion times from the constant anatomical structure. However, this approach critically depends on accurately separating the anatomical structure from the contrast information. Additionally, registration occurs solely between the isolated anatomical images, neglecting the signal model along the inversion time axis. As a result, this method may lead to physically unrealistic deformations due to its lack of integration with the underlying signal model.

However, these approaches do not explicitly incorporate the signal model into their registration systems, despite its critical role in governing image contrast changes. Consequently, they may produce physically unrealistic deformations. In contrast, classical model-based methods account for these issues, but their iterative nature renders them comparatively slow.

Previously, we introduced PCMC-T1, a method designed explicitly for free-breathing myocardial T1 mapping that incorporates physical constraints into the motion correction process \cite{hanania2023pcmc}. This deep learning-based approach uses a supervised framework to encourage physically realistic cardiac T1 mapping under free-breathing conditions. However, the practical application of PCMC-T1 is limited by its reliance on large annotated datasets and extensive training, which are not readily available for every scanner and protocol.

This work introduces MBSS-T1, a subject-specific self-supervised approach extending our previous PCMC-T1 model. The proposed MBSS-T1 approach enables the generation of motion robust T1 maps across various imaging protocols, eliminating the need for protocol-specific data collection, manual annotations, and extensive training. 

\section{Methods}

The T1 recovery time can be modeled using distinct equations tailored to the specific sequences employed for signal measurement. The overall signal recovery for inversion recovery and Look-Locker-based measurements can generally be described as follows:
\begin{equation}
I(t) = A-B\cdot e^{-t/T1^*}   
\label{eq.recovery_equation_molli}
\end{equation}
where \( I(t) \) is the magnitude image reconstructed from the multi-coil raw data obtained at time \( t \). 

For the modified Look-Locker inversion recovery (MOLLI) sequence, $A$ and $B$ are fitting parameters related to the equilibrium magnetization and type of preparation, $t$ is the time after the preparation (i.e. inversion), and $T1^*$ is the apparent $T1$ which needs to be further corrected to get the actual tissue $T1$ using the following equation \cite{taylor2016t1}: 
\begin{equation}
T1 = T1^*\cdot(B/A-1)
\label{eq.molli_correction}
\end{equation}
For the slice-interleaved T1 (STONE) sequence, the two-parameter signal recovery model (i.e. assuming a perfect efficiency of the inversion pulses) can be derived from Eq.~\ref{eq.recovery_equation_molli} as follows: $A$ can be substituted by the equilibrium magnetization $M_0$ which is the magnetization of the tissue before any preparation occurs, $B$ by $2M_0$, and $T1^*$ by the actual $T1$ to get the proper signal recovery equation  for this sequence  \cite{weingartner2015free}:
\begin{equation}
I(t) = M_0 \cdot \left(1 - 2 \cdot e^{-t/T1}\right)\,
\label{eq.recovery_equation_stone}
\end{equation}

The estimation of the signal recovery model parameters $\Theta$ (i.e. $A, B, T1^*$ for the MOLLI sequence and $M_0, T1$ for the STONE sequence) is typically performed by solving a least-squares optimization problem \cite{taylor2016t1}. This involves minimizing the sum of the squared differences between the measured magnitude images and the values predicted by the model as follows:
\begin{equation}
\widehat{\Theta} = \argmin_{\Theta} \sum_{i=0}^{N-1} \left( f\left(\Theta, t_i\right) - I_i \right)^2
\end{equation}
where $I_i$ represents the observed magnitude images at different time points \( t_i \), \( N \) is the total number of measurements, $\Theta$ are the model parameters. 

This classical least-squares approach, as well as deep learning methods such as the one described in \cite{guo2022accelerated}, assumes no motion between the different images, typically achieved through prior image registration techniques. However, achieving precise registration is particularly challenging due to the inherent complexities of the image data, such as contrast inversion, partial volume effects, and signal nulling for images acquired near the zero-crossing of the T1 relaxation curve. Furthermore, sequential registration and model fitting methods may converge to local minima, leading to pairwise registrations that fail to exhibit physically accurate signal recovery. To explicitly address motion during model fitting, the optimization problem must be expanded to incorporate deformation fields, which can be treated as additional variables within the least-squares framework as follows: 
\begin{equation}
\widehat{\Theta}, \widehat{\Phi} = \argmin_{\Theta, \Phi} \sum_{i=0}^{N-1} \left\| f(\Theta, t_i) - \phi_i \circ I_i \right\|^2\
\end{equation}
In this extended formulation, $\widehat{\Theta}$ are the estimated signal recovery parameters (Eq.~\ref{eq.recovery_equation_molli}, or Eq.~\ref{eq.recovery_equation_stone} according to the sequence used to acquire the data), while $\widehat{\Phi}$ represents the estimated deformation fields that account for motion. Specifically, $\phi_i$ denotes the $i$-th deformation field between the $i$-th image and the reference image $I_r$, with $\Phi=\{\phi_i\}_{i=0}^{N-1}$ and $I_i$ corresponds to the $i$-th original magnitude image, and $\phi_i \circ I_i$ is the deformed image, reflecting the influence of motion on the observed magnetization values. In case $I_i$ is the reference image itself, $\phi_i$ is the identity deformation field.

Direct optimization of this equation is challenging due to the high dimensionality of the problem, stemming from the numerous unknowns associated with the signal recovery model and the deformation fields. While iterative solutions using classical methods, such as those outlined in \cite{xue2012motion, tilborghs2019robust}, are feasible, they are often computationally expensive and susceptible to convergence at local minima.

To overcome these limitations, we reframe the optimization task as estimating the weights of a deep neural network ($\Psi$) that simultaneously predicts the signal recovery model parameters and spatial transformations from the observed magnitude images at the corresponding time points. This is achieved using two encoder-decoder modules that operate concurrently: one for predicting the signal recovery model parameters (parametric mapping component) and one for predicting the deformation fields (motion component). The objective function to be minimized is expressed as:
\begin{equation}
\Psi = \argmin_{\Psi} \sum_{i=0}^{N-1} \left\|f\left(NNmap_{\Psi},t_i\right)  -  NNreg_{\Psi}(i) \circ I_i  \right\|^2
\end{equation}
where $NNmap_{\Psi}$ represents the output of the parametric mapping component, $f\left(NNmap_{\Psi},t_i\right)$ are the magnitude images synthesized from the estimated parameters at the time $t_i$ using the signal recovery model (Eq.~\ref{eq.recovery_equation_molli}, or Eq.~\ref{eq.recovery_equation_stone} according to the sequence used to acquire the data), and $NNreg_{\Psi}(i)$ denotes the output of the motion component, which includes the predicted deformation fields between the images $I_i$ and the reference image $I_r$.

\subsection{Model Architecture}
\begin{figure*}[t!]\includegraphics[width=0.95\textwidth]{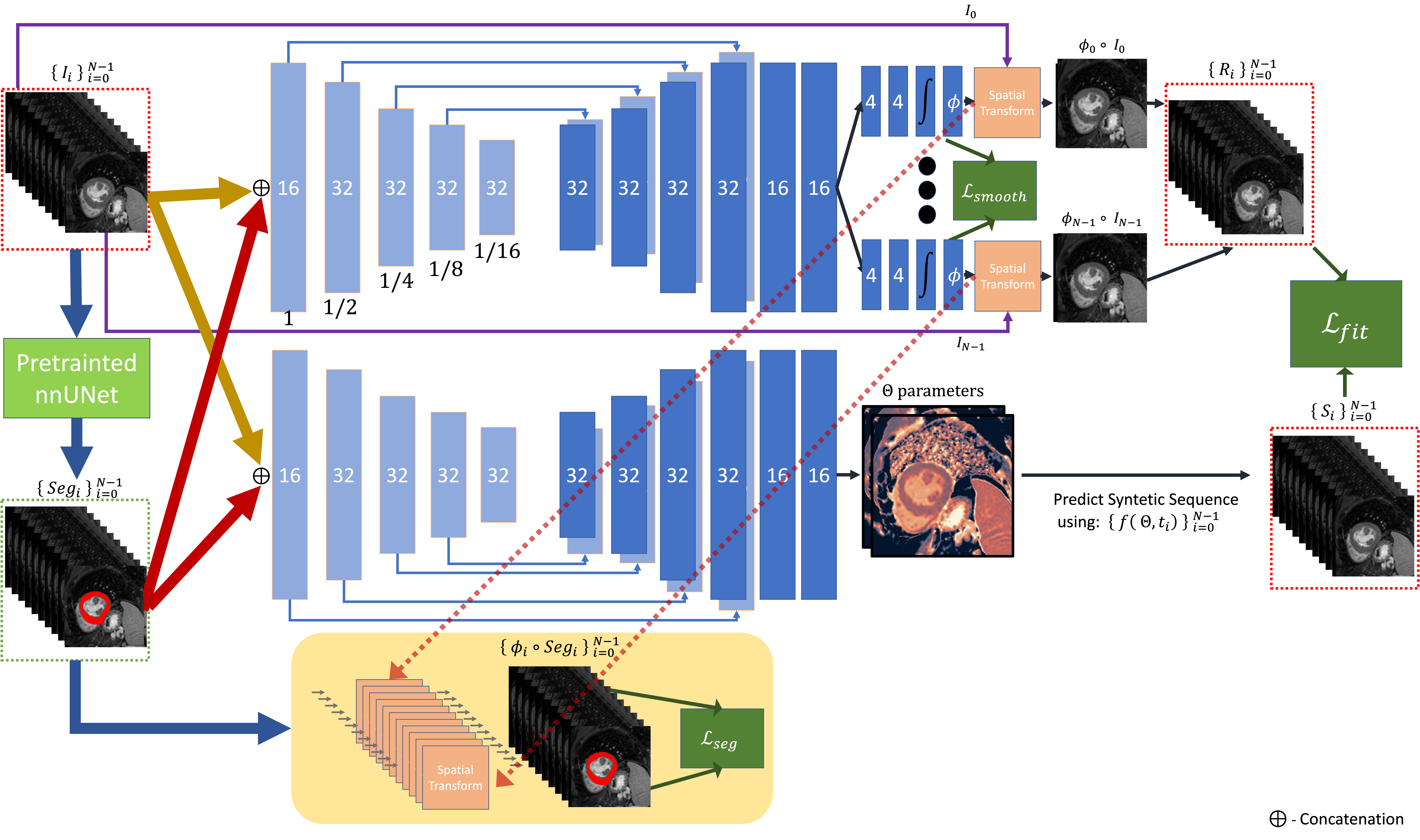}
\caption{Overview of the MBSS-T1 model architecture: (a) The motion component is responsible for deformable image registration, predicting deformation fields that align the acquired images across different time points. (b) The parametric mapping component estimates the signal recovery model parameters and generates synthetic aligned T1-weighted (T1W) images. (c) The segmentation and the confidence component uses a pre-trained segmentation network to extract the myocardium segmentations, calculates the deformed segmentations, and retains the ones with high confidence. The network aims to minimize the distance between the motion-corrected T1W and synthetic images while maximizing the Dice score between the deformed segmentations.}

\label{fig:network_architecture}
\end{figure*}

Figure~\ref{fig:network_architecture} presents our overall architecture. Our system comprises a pre-trained nnUNet for myocardium segmentation and two U-Net-inspired encoder-decoder modules that operate simultaneously to predict the deformation fields and the signal recovery model parameters. We describe each component in detail below. 

 \textbf{Segmentation and Confidence Component:}  First, an nnUNet segmentation network \cite{isensee2021nnu} pre-trained on the STONE dataset \cite{el2018nonrigid,DHEUAV_2019} is utilized to predict the myocardium segmentation for each frame. This network was trained using a 5-fold cross-validation setup to avoid data leakage. The output of the nnUNet includes both the segmentation of the myocardium and a pixelwise confidence score for the segmentation. Next, a confidence block is employed to retain only the segmentations with high confidence scores. This block receives the masks and confidence maps generated by the nnU-Net output. It retains only those masks that meet two criteria: (1) a required percentage \( \gamma \) of the myocardial segmentation pixels must have a score greater than a confidence threshold \( \alpha \), and (2) the segmentation mask must consist of exactly one connected component. Here, \( P_i(x) \) denotes the softmax probability at pixel \( x \) in the \( i \)-th segmentation, and \( S_i \) is the corresponding segmentation mask. Let \( G \) be the set of indices of segmentations that satisfy both of these criteria:
\begin{equation}
 G = \left\{ i \mid \frac{1}{|S_i|} \sum_{x \in S_i} \mathbf{1}\left[P_i(x) > \alpha\right] \geq \gamma \text{ and } S_i \text{ is connected} \right\}
\end{equation}
Next, we define a reference image index \( r \) as the time point corresponding to the segmentation with the highest confidence score:
\begin{equation}
r = \underset{i \in G}{\arg\max} \left( \frac{1}{|S_i|} \sum_{x \in S_i} P_i(x) \right)
\end{equation}
This equation selects the index \( r \) where the average confidence across the segmentation \( S_i \) is the highest among the segmentations in the set \( G \). This reference image will serve as the fixed image for the registration network. In addition, we define the extended myocardium mask \( K \) as the bitwise OR of the segmentations in the set \( G \).
 
\textbf{Motion Estimation Component:} The upper module, an extension of the VoxelMorph architecture, functions as a multi-image deformable image registration module. It dynamically adjusts the number of registration heads based on the number of the magnitude images  ($N$) reconstructed from the acquired data. Its input consists of a concatenation of the reference image $I_r$, selected as the frame with the highest confidence score from the segmentation and confidence component output, the remaining images $I_i, i \neq r$ as the moving images, and the segmentations retained in the set $G$, which includes only those with high confidence. The output of this module is a set of $N-1$ deformation fields $\{\phi_i\}, i\neq r$ between the moving images $I_i, i\neq r$ and the reference image $I_r$ as well as the deformed images $\phi_i\circ I_i, i\neq r$ for each time point $i$. This module leverages the deformation field prediction layers from \cite{balakrishnan2019voxelmorph}, and the integration layer from \cite{dalca2019unsupervised} to encourage diffeomorphism. 

Next, the segmentation masks associated with each time point are aligned using the corresponding deformation fields to obtain deformed segmentations. 

\textbf{Parametric Mapping Component:} The bottom module takes the same input as the Motion Estimation Component and predicts the parameters of the signal recovery model, forwarding these parameters to the signal generation layer. This layer predicts a sequence of synthetic images \(\{s_i \mid i=0, \ldots, N-1\}\) following the signal recovery equation suitable for the sequence used for the acquisition (i.e. Eq.~\ref{eq.recovery_equation_molli} for the MOLLI sequence and Eq.~\ref{eq.recovery_equation_stone} for the STONE sequence). These images are anatomically aligned according to the correct signal recovery model, as each image prediction utilizes the same signal recovery model parameters, with the only difference being the times point of each image.

\subsection{Loss Functions}
We encourage predictions of physically and anatomically plausible deformation fields by coupling three terms in our loss function as follows: 
\begin{equation}
\mathcal{L}_{total} = \lambda_1\cdot \mathcal{L}_{fit} + \lambda_2\cdot \mathcal{L}_{smooth} + \lambda_3\cdot \mathcal{L}_{Seg} 
\end{equation}
The first term ($\mathcal{L}_{fit}$) addresses the differences between the images synthesized by the signal recovery layer equation (i.e. Eq.~\ref{eq.recovery_equation_molli} for the MOLLI sequence and Eq.~\ref{eq.recovery_equation_stone} for the STONE sequence) from the parameters predicted by the parametric mapping component and the deformed images. To achieve this, we compute the mean squared error (MSE) between the registered images ${R_i | i=0, \ldots, N-1}$ and the synthetic images ${S_i | i=0, \ldots, N-1}$ after applying the mask \( K \):
\begin{equation}
\begin{split}
\mathcal{L}_{fit}\left(\Theta, t_{i=0}^{N-1}, \Phi\right) = & \sum_{i=0}^{N-1}\left((S_i - R_i)\cdot K\right)^2 \\
= & \sum_{i=0}^{N-1} \left((f\left(\Theta,t_i\right) - \phi_i \circ I_i)\cdot K \right)^2
\end{split}
\end{equation}
Here, $S_i$ denotes the images generated using the signal recovery equation (Eq.~\ref{eq.recovery_equation_molli} or Eq.~\ref{eq.recovery_equation_stone}) using the predicted model parameters ($\Theta$), while the registered images are obtained by applying the deformation fields predicted by the motion estimation component $\Phi$ to the acquired magnitude images $I_i$. This term enforces the creation of deformation fields that are physically plausible by ensuring consistency with the T1 signal recovery model.

The second term ($\mathcal{L}_{smooth}$) encourages the creation of realistic and smooth deformation fields $\Phi$. This is achieved by penalizing large $l_2$ norms of the gradients of the velocity fields \cite{balakrishnan2019voxelmorph}:
\begin{equation}
\mathcal{L}_{smooth}(\Phi) = \sum_{i=0}^{N-1} \frac{1}{\Omega} \sum_{p \in \Omega} \norm{\nabla \phi_i(p)}^2\,.
\end{equation}
Here, $\Omega$ denotes the domain of the velocity field, and $p$ represents the voxel locations. By minimizing this term, the model promotes smoother deformation fields, reducing unrealistic distortions.

The last term (\(\mathcal{L}_{Seg}\)), inspired by \cite{balakrishnan2019voxelmorph}, encourages anatomically correct deformation fields by maximizing the Dice score between the deformed myocardium segmentations and by maximizing the Dice score between the deformed left ventricle segmentations. We denote \(\mathcal{L}_{Seg}\) as follows:
\begin{equation}
\mathcal{L}_{Seg} = \mathcal{L}_{MyoSeg} + \mathcal{L}_{LVSeg}
\end{equation}

Our approach focuses on the segmentations retained in the set \( G \), which contains only those segmentations with high confidence. The \(\mathcal{L}_{MyoSeg}\) and \(\mathcal{L}_{LVSeg}\) segmentation loss functions are therefore defined using only these high-confidence segmentations:
\begin{equation}
\begin{split}
\mathcal{L}_{\text{MyoSeg}}&(r, \{MyoSeg_i\}_{i \in G}, \{\phi_i\}_{i \in G}) = \\ 
& \sum_{\substack{i \in G \\ i \neq r}} \text{DiceLoss}(MyoSeg_r, MyoSeg_i \circ \phi_i)
\end{split}
\end{equation}

\begin{equation}
\begin{split}
\mathcal{L}_{\text{LVSeg}}&(r, \{LVSeg_i\}_{i \in G}, \{\phi_i\}_{i \in G}) = \\
& \sum_{\substack{i \in G \\ i \neq r}} \text{DiceLoss}(LVSeg_r, LVSeg_i \circ \phi_i)
\end{split}
\end{equation}
where \( MyoSeg_i \) is the \( i \)-th binary segmentation mask of the myocardium, \( LVSeg_i \) is the \( i \)-th binary segmentation mask of the left ventricle, and \( Seg_r \) is the binary segmentation mask of the \( r \)-th image, which is the reference image. Only the segmentations indexed by \( G \), which meet the high-confidence criteria, are used to calculate the loss.

\subsection{Implementation details}
Our network architecture includes two UNet backbones. Each backbone employs 2D convolutions in both the encoder and decoder stages, using a kernel size of 3, stride of 1, and padding of 1. The encoder consists of three convolutional layers with [16, 32, 32] filters, each followed by a LeakyReLU activation function with a negative slope of 0.2. To progressively reduce spatial dimensions, a max-pooling operation with a kernel size of 2, stride of 2, padding of 0, and dilation of 1 is applied after each convolutional layer.

The decoder contains four convolutional layers with 32 filters, each followed by a LeakyReLU activation with a slope of 0.2, along with an upsampling operation with a scaling factor of 2. Subsequently, three additional convolutional layers with [32, 16, 16] filters are applied, each followed by a LeakyReLU activation function.

In the registration heads, each head comprises two convolutional layers with [4, 4] filters, each followed by a LeakyReLU activation with a slope of 0.2. This is followed by another convolutional layer with 2 filters, and a resize block with a scaling factor of 0.5. The velocity field is predicted at every two pixels and processed through seven integration steps, followed by a resizing operation with a factor of 2.

In the parameter heads, each head consists of two convolutional layers with a single filter, each followed by a LeakyReLU activation with a slope of 0.2. This setup ensures the integration of both registration and parameter estimation components into the network's architecture.

We implemented our models in PyTorch. Experimentally, we zeroed the \( r \)-th deformation field and predicted deformation fields only for the rest of the time points. We optimized our hyperparameters using a grid search. The final settings for the loss function parameters were: \(\lambda_1 = 1\), \(\lambda_2 = 5000\), \(\lambda_3 = 80000\), \(\gamma = 0.99\), and \(\alpha = 0.9\). We used a batch size of 1 and the ADAM optimizer with a learning rate of \(2 \cdot 10^{-3}\). Our network overfits each case. For each case, we trained the network from scratch for a fixed number of 40 iterations, which took approximately 6 seconds. We performed hyperparameter optimization for the baseline methods using a grid search. All experiments were run on an NVIDIA Tesla V100 GPU with 32GB of RAM.

\section{Evaluation Methodology}
\subsection{Datasets}
\subsubsection{Free Breathing - STONE Dataset}
A publicly available myocardial T1 mapping dataset \cite{el2018nonrigid, DHEUAV_2019} consists of 210 subjects, 134 males and 76 females, aged $57\pm14$ years. All subjects were diagnosed with or suspected of having cardiovascular diseases. The imaging was performed on a 1.5T MRI scanner (Philips Achieva) equipped with a 32-channel cardiac coil, using an ECG-triggered, free-breathing, respiratory-navigated, slice-interleaved cardiac T1 mapping sequence (STONE) \cite{weingartner2015free}. The acquisition parameters were as follows: field of view (FOV) $=360\times351$ [mm\(^2\)] and voxel size $=2.1\times2.1\times8$ [mm\(^3\)]. Five slices were captured from the base to the apex in the short-axis view for each subject at 11 distinct time points. Additionally, the dataset included manual expert segmentations of the myocardium \cite{el2018nonrigid}. The images were resized to $160\times160$ pixels per time point, and min-max normalization was applied to the entire sequence to standardize the image intensities.

\subsubsection{Free Breathing and Breath Hold - MOLLI Dataset}
Data was acquired on a 3T Siemens MRI system (PRISMA, Siemens Healthineers, Erlangen, Germany) equipped with a 32-channel body coil, using an ECG-triggered, slice-interleaved cardiac T1 mapping MOLLI sequence (MyoMaps, SIEMENS Healthcare). The acquisition parameters were as follows: field of view (FOV) $=306\times360$ [mm\(^2\)], flip angle of 35$^\circ$, and slice thickness of 8 mm. Five slices were captured from the base to the apex in the short-axis view for each subject at eight distinct time points. The dataset consists of cardiac images obtained from 19 patients undergoing two scans: one during breath-hold and another during free breathing. The data was acquired without any respiratory navigation. Additionally, extra breath-hold and free-breathing scans were performed for five patients for test-retest analysis. Pre-processing was performed following the same procedure described above for the STONE dataset.

\subsection{Experiments}

\subsubsection{Exp. 1: Motion Correction in Free-Breathing MRI Using STONE Dataset}

\textbf{Experimental setup:} We used an MBSS-T1 version (MBSS-T1$_\textrm{STONE}$) with a suitable model parameters estimation module (i.e Eq.~\ref{eq.recovery_equation_stone}). We compared our MBSS-T1$_\textrm{STONE}$ against two variations: a multi-image registration model using a mutual-information-based loss function (REG-MI) \cite{hanania2023groupT1}, and PCMC-T1, which incorporates physical constraints into the motion correction process within an unsupervised framework \cite{hanania2023pcmc}.
Additionally, two state-of-the-art deep learning algorithms for medical image registration were included in the comparison: pairwise probabilistic diffeomorphic VoxelMorph with a mutual-information-based loss \cite{dalca2019unsupervised} and pairwise SynthMorph \cite{hoffmann2021synthmorph}.

For the unsupervised approaches (PCMC-T1, REG-MI, VoxelMorph) and the nnUNet pretrained segmentation network (used for MBSS-T1$_\textrm{STONE}$), a 5-fold cross-validation strategy was implemented. In each fold, the 210 subjects were divided into 80\% for training and 20\% for testing. For each fold, the VoxelMorph model was trained in an unsupervised manner, with the fixed and moving images randomly selected from the training data. The reported results represent the aggregation of the test data across the entire dataset.

The SynthMorph training procedure does not require any training data and was therefore applied directly to the entire STONE dataset. Similarly, MBSS-T1$_\textrm{STONE}$ was applied directly to the entire STONE dataset without any prior training.

The reference time point for registration was automatically determined by the MBSS-T1 approach and was subsequently set to the first time point for all other methods.

T1 maps generated directly from the acquired images were also used for comparison by applying a standard cardiac T1 mapping procedure. A T1 relaxation curve was fitted to the MRI data using the conventional inversion recovery model without any pre-motion correction techniques. Fitting was performed per pixel across the series of T1-weighted images, resulting in maps representing the T1 relaxation times. These maps, referred to as 'w.o motion correction,' serve as a baseline to evaluate the impact of motion correction on the accuracy and quality of the cardiac T1 mapping.

The T1 maps generated by our MBSS-T1$_\textrm{STONE}$ model were quantitatively evaluated against those produced both before (`w.o motion correction') and after deep-learning-based image registration as a pre-processing step. The evaluation metrics used were the $R^2$ of the model fit to the observed data in the myocardium, the Dice score, and the Hausdorff distance values of the myocardium segmentations. In addition, we evaluated the capacity of the different methods to generate diffeomorphic deformation fields by assessing the determinants of the deformation fields Jacobians for each method.

\textbf{Clinical impact:} The clinical implications of our method were assessed through a semi-quantitative ranking of the T1 maps for motion artifacts. This assessment was conducted by an expert cardiac radiologist with five years of experience (I.V), blinded to the methods used to generate the maps. A random selection of 29 cases (each with five slices) from the test set and their corresponding T1 maps was evaluated. The radiologist rated each slice, assigning a score of 1 to slices without significant visible motion artifacts and a score of 0 to those with artifacts. Overall patient scores were calculated by summing the scores of the slices. The highest possible score per subject was 5, indicating no motion artifacts in all slices, while the lowest was 0, indicating artifacts in all slices. Statistical significance was evaluated using the repeated measures ANOVA test, with p$<$0.05 considered significant.

\subsubsection{Exp. 2: Motion Correction in Free-Breathing and Breath-Hold MRI Using MOLLI}

\textbf{Experimental Setup:} The objective of this experiment was to evaluate the effectiveness of the MBSS-T1 method in accounting for motion during free-breathing and breath-hold MRI scans using the MOLLI sequence. We utilized the pre-trained nnUNet model trained on the STONE dataset, as described in the previous section. To make the MBSS-T1 model compatible with the MOLLI sequence (MBSS-T1$_\textrm{MOLLI}$), we modified the MBSS-T1 model by adjusting the number of heads in the network output from 11 to 8 and replacing the model fitting layers with layers that match the MOLLI sequence signal recovery model and T1 correction equation(Eq.~\ref{eq.recovery_equation_molli}, ~\ref{eq.molli_correction}). The performance of four motion-correction methods was compared: (1) without motion correction, (2) Siemens MyoMaps with motion correction \cite{xue2012motion}, (3) Synthmorph \cite{hoffmann2021synthmorph} and (4) MBSS-T1$_\textrm{MOLLI}$. 

The reference time point for registration was automatically determined by the MBSS-T1 approach and was subsequently set to the first time point for Synthmorph.

For Siemens MyoMaps, we used the motion-corrected T1 maps produced by the scanner's image processing software. We quantitatively evaluate our method against comparative approaches using three key metrics: the coefficient of determination (\(R^2\)) for model fit to the observed myocardial data, the Dice score, and the Hausdorff distance for myocardial segmentations. Since manual myocardial annotations are unavailable for our MOLLI dataset, we applied a pretrained segmentation network to each frame individually and computed the segmentation metrics both before and after motion correction. These metrics were not measured for Siemens MyoMap, as we do not have access to the registered T1-weighted images.

\textbf{Clinical Impact:} As in Exp. 1, the clinical impact of our method was assessed through a semi-quantitative ranking of the T1 maps for the presence of motion artifacts. This assessment was conducted on free-breathing and breath-hold scans to determine how effectively each method mitigated motion artifacts. The primary evaluation metric was a clinical score assigned by an experienced cardiac radiologist with five years of experience based on the clarity and quality of the T1 maps produced by each method.

\subsubsection{Exp. 3: Consistency of cardiac T1 Mapping with Motion Correction in Test/ReTest MOLLI Dataset}

\textbf{Experimental Setup and Objective:} 
This experiment aimed to assess the consistency and reproducibility of cardiac T1 mapping when motion correction is applied. This analysis aimed to evaluate whether motion correction techniques improve the reliability of myocardial T1 measurements across repeated scans, particularly in the presence of motion. The subset included data from five patients undergoing two free-breathing and two breath-hold scans. It is important to note that the test and retest images are not aligned, as the subject left the room for a 30-minute break between scans. As a result, the images are not directly comparable on a pixel level. Instead, comparisons between the test and retest data are conducted at a higher level of abstraction, focusing on regional metrics rather than pixel-level correspondence.

\textbf{Quantitative Evaluation:} 
To quantitatively assess the reproducibility of T1 measurements, we calculated the intraclass correlation coefficient (ICC3) \cite{mcgraw1996forming} for mean myocardial T1 values across three different methods: without motion correction, Siemens MyoMaps, and MBSS-T1. The analysis was conducted using data from five patients in the MOLLI dataset, each of whom underwent two free-breathing and two breath-hold scans. For each scan and method, mean myocardial T1 values were computed for AHA-16 segments. The ICC3 metric was then calculated for all paired test-retest segments to assess the agreement between repeated measurements. A high ICC3 value indicates strong agreement between the test and retest measurements, suggesting robustness and reproducibility in quantifying myocardial T1 values. This metric provides valuable insights into the reliability and consistency of T1 measurements across different methods and scanning conditions. A comprehensive understanding of each method's performance in clinical practice can be achieved by quantifying the agreement between test and retest measurements across different ROIs and scanning conditions. This analysis also enables the evaluation of the effectiveness of motion correction techniques in improving measurement accuracy and reproducibility.

\section{Results}

\begin{figure}[t!]\centering\includegraphics[width=0.95\columnwidth]{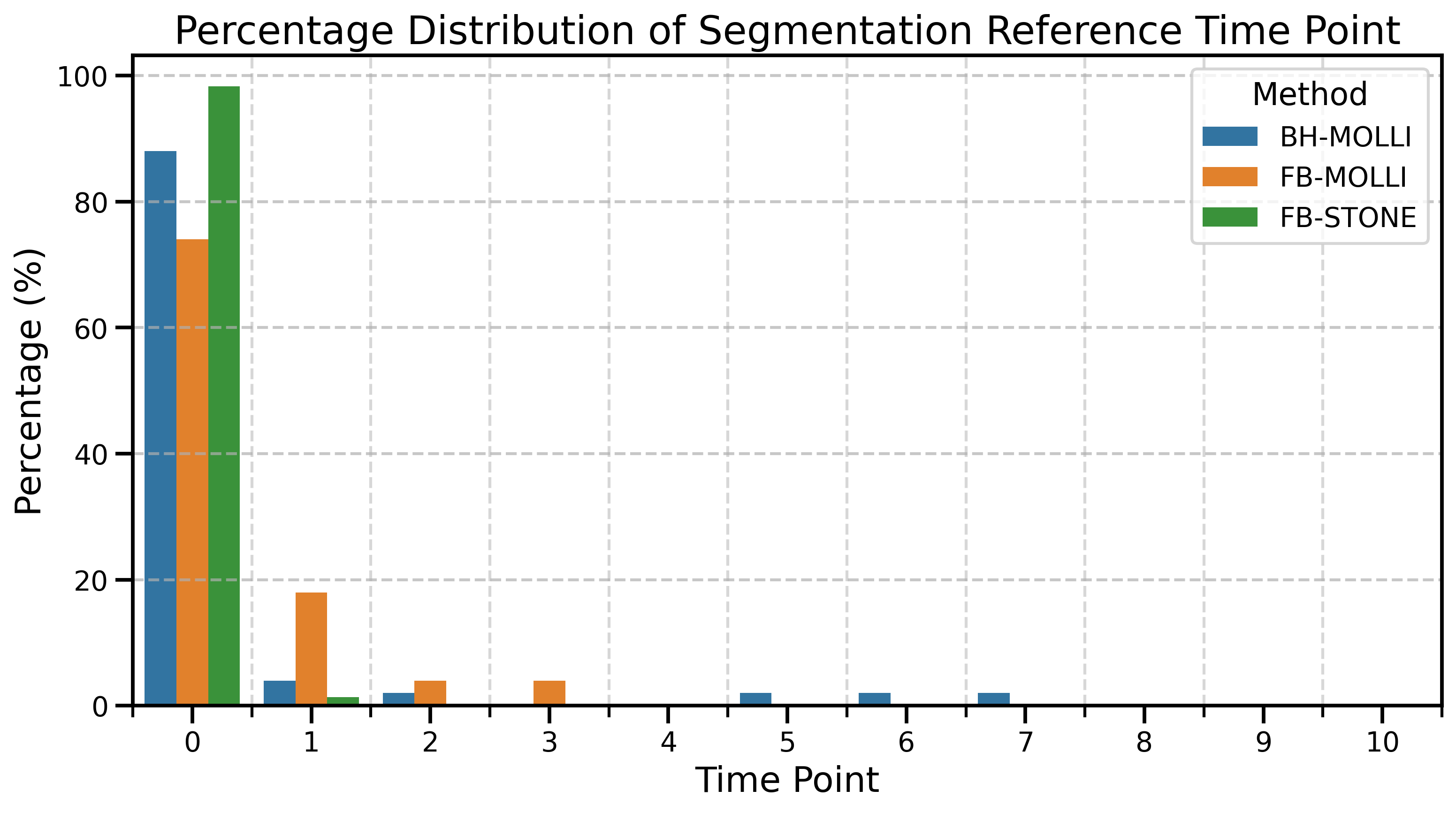}
\caption{The distribution of the selected reference time points across different acquisition methods (BH-MOLLI, FB-MOLLI, and FB-STONE) is presented in the figure below. The x-axis represents the time points within the respective sequences, while the y-axis indicates the percentage of cases in which each time point was chosen. The percentage for MOLLI (BH-MOLLI, FB-MOLLI) is calculated based on a sequence length of 8 time points, whereas for STONE (FB-STONE), the sequence length is 11 time points. }
\label{fig:ref_ind_time_point_distribution}
\end{figure}

Figure \ref{fig:ref_ind_time_point_distribution} presents the distribution of the selected reference time points across different acquisition methods (BH-MOLLI, FB-MOLLI, and FB-STONE) by our MBSS-T1 approach. The majority of selections occur at time point 0 across all methods (BH-MOLLI, FB-MOLLI, and FB-STONE). These findings suggest that the reference state is predominantly chosen in the mid-diastolic phase.

\subsection{Exp. 1: Motion Correction in Free-Breathing MRI Using STONE Dataset}

{\bf Quantitative evaluation:} Table~\ref{table:results_stone} summarizes our results for the test sets across all folds, encompassing a total of 210 patients. The MBSS-T1$_\textrm{STONE}$ method outperformed other state-of-the-art registration methods in terms of \(R^2\), Dice score, and Hausdorff distance. Specifically, MBSS-T1$_\textrm{STONE}$ achieved an \(R^2\) value of \(0.975 \pm 0.05\), which represents an improvement of approximately 3\% compared to the next best method, PCMC-T1, with an \(R^2\) of \(0.955 \pm 0.078\). Additionally, MBSS-T1$_\textrm{STONE}$ produced a Dice score of \(0.89 \pm 0.075\), surpassing SynthMorph's \(0.88 \pm 0.149\), and reduced the Hausdorff distance to \(6.43 \pm 5.54\) mm, which is a 31\% improvement compared to SynthMorph's \(8.59 \pm 9.98\) mm. These results suggest that our approach enhances the physical plausibility of deformations through signal relaxation and anatomical consistency. Our results are particularly noteworthy, considering we compared methods requiring a full training process.

\begin{table*}[t]
\centering
\caption{Quantitative comparison between motion correction methods for STONE myocardial T1 mapping.
Results are presented as mean$\pm$std.}
\footnotesize
\begin{tabular}{|c|c|c|c|c|}
\hline
\makecell{} & \makecell{$R^2 \uparrow$} & \makecell{$DSC \uparrow$} & \makecell{HD [mm] $\downarrow$} & \makecell{Clinical Score $\uparrow$} \\\hline
w.o motion correction & $0.911\pm0.12$ & $0.664\pm0.23$ & $14.93\pm11.76$ & $2.9\pm0.92$ \\
SynthMorph & {$0.946\pm0.09$} & {$0.88\pm0.149$} & {$8.59\pm9.98$} & {$3.66\pm0.83$} \\
Voxelmorph-seg & $0.941\pm0.096$ & $0.84\pm0.188$ & $9.39\pm11.93$ & $3.38\pm0.65$  \\
Reg-MI & $0.95\pm0.08$ & $0.73\pm0.168$ & $16.29\pm11.43$ & {$3.68\pm0.83$}  \\
\hline 
PCMC-T1 & {$0.955\pm0.078$} & $0.835\pm0.137$ & {$9.34\pm7.85$} & {$3.93\pm0.78$} \\
MBSS-T1$_\textrm{STONE}$ & \textbf{0.975$\pm$0.05} & \textbf{0.89$\pm$0.075} & \textbf{6.43$\pm$5.54} & \textbf{4.33$\pm$0.54} \\
\hline
\end{tabular}
\label{table:results_stone}
\end{table*}

\begin{figure*}[t!]\includegraphics[width=1\textwidth]{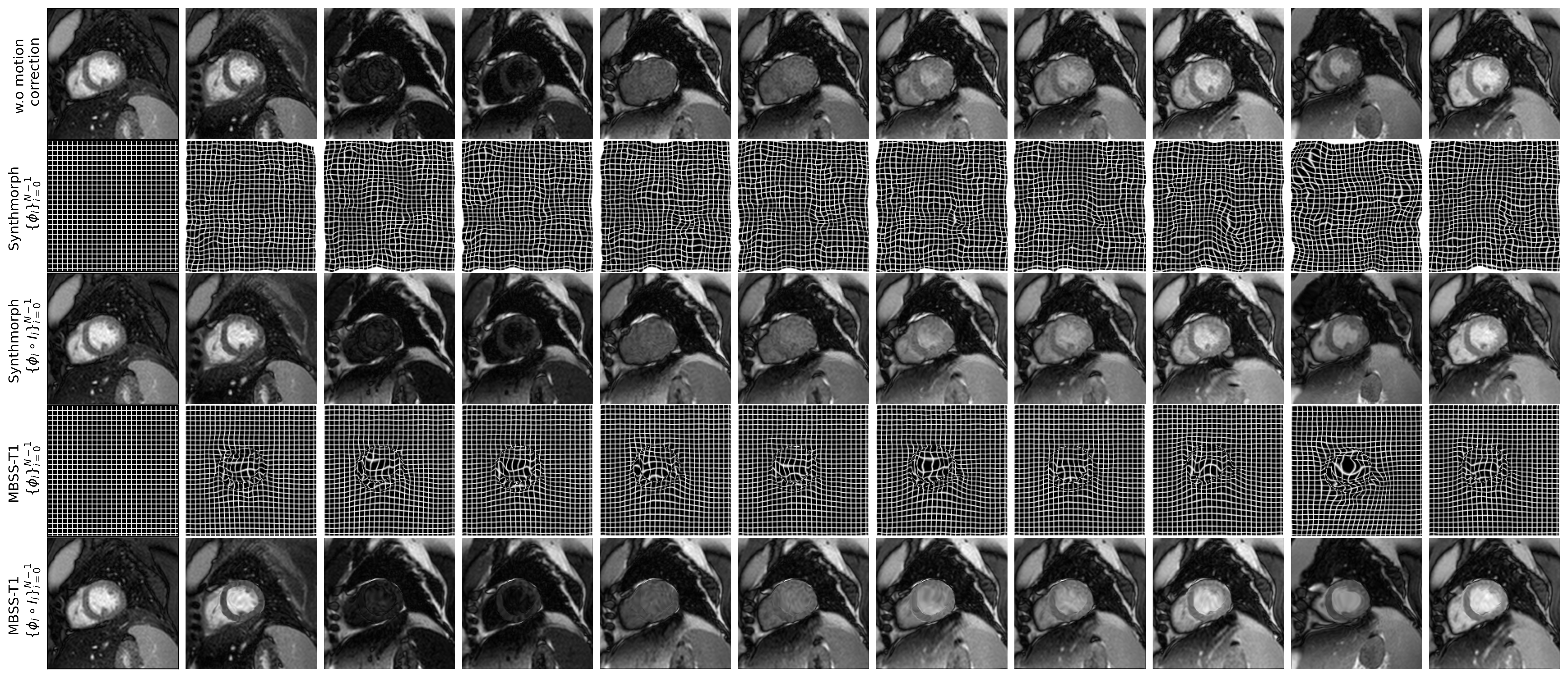}
\centering
\caption{Comparison of deformation grids and registered images using different motion correction methods across multiple frames. The top row illustrates images without motion correction, followed by deformation grids and registered images produced by SynthMorph. The third row presents results from the MBSS-T1$_\textrm{STONE}$ method, including the deformation grids and corresponding registered images. }
\label{fig:grids}
\end{figure*}

Figure~\ref{fig:grids} depicts the deformed images alongside the predicted deformation fields for the different methods. Our MBSS-T1 approach generates smoother grids with visually realistic deformations, whereas SynthMorph produces less realistic deformation patterns.

Table~\ref{table:jacobian_comparison} summarizes the results of the Jacobian determinant analysis. Our MBSS-T1 approach produces deformation fields with a low mean number of folds ($0.0052 \pm 0.111$) and a mean Jacobian determinant close to 1 ($0.9969 \pm 0.0043$), indicating more realistic deformation fields.

\begin{table}[t]
\centering
\caption{Quantitative comparison of registration methods for the STONE myocardial T1 mapping dataset using the mean Jacobian determinant (\(|J_\phi|\)) and the mean number of locations with non-positive Jacobian determinants (\(|J_\phi \leq 0|\)) for each registration field. \(|J_\phi|\) values close to 1 indicate smooth deformations and stable transformations, while lower values of \(|J_\phi \leq 0|\) reflect fewer distortions and better registration quality.}
\footnotesize
\begin{tabular}{|c|c|c|c|c|}
\hline
\makecell{Method} & \makecell{Mean $|J_\phi|$} & \makecell{$|J_\phi \leq 0|$} \\\hline
w.o motion correction & $1$ & $0$\\ 
Voxelmorph & $0.9909\pm0.0134$ & $0.0000\pm0.0000$\\
SynthMorph & $0.9978\pm0.0118$ & $0.0009\pm0.0039$ \\
RegMI & $1.0231\pm0.0226$ & $2476.1762\pm1443.7345$\\
\hline
PCMC & $1.0078\pm0.0049$ & $82.3292\pm77.1942$ \\
MBSS-T1$_\textrm{STONE}$ & $0.9969\pm0.0043$ & $0.0052\pm0.111$ \\
\hline
\end{tabular}
\label{table:jacobian_comparison}
\end{table}

\begin{figure*}[t!]\includegraphics[width=0.95\textwidth]{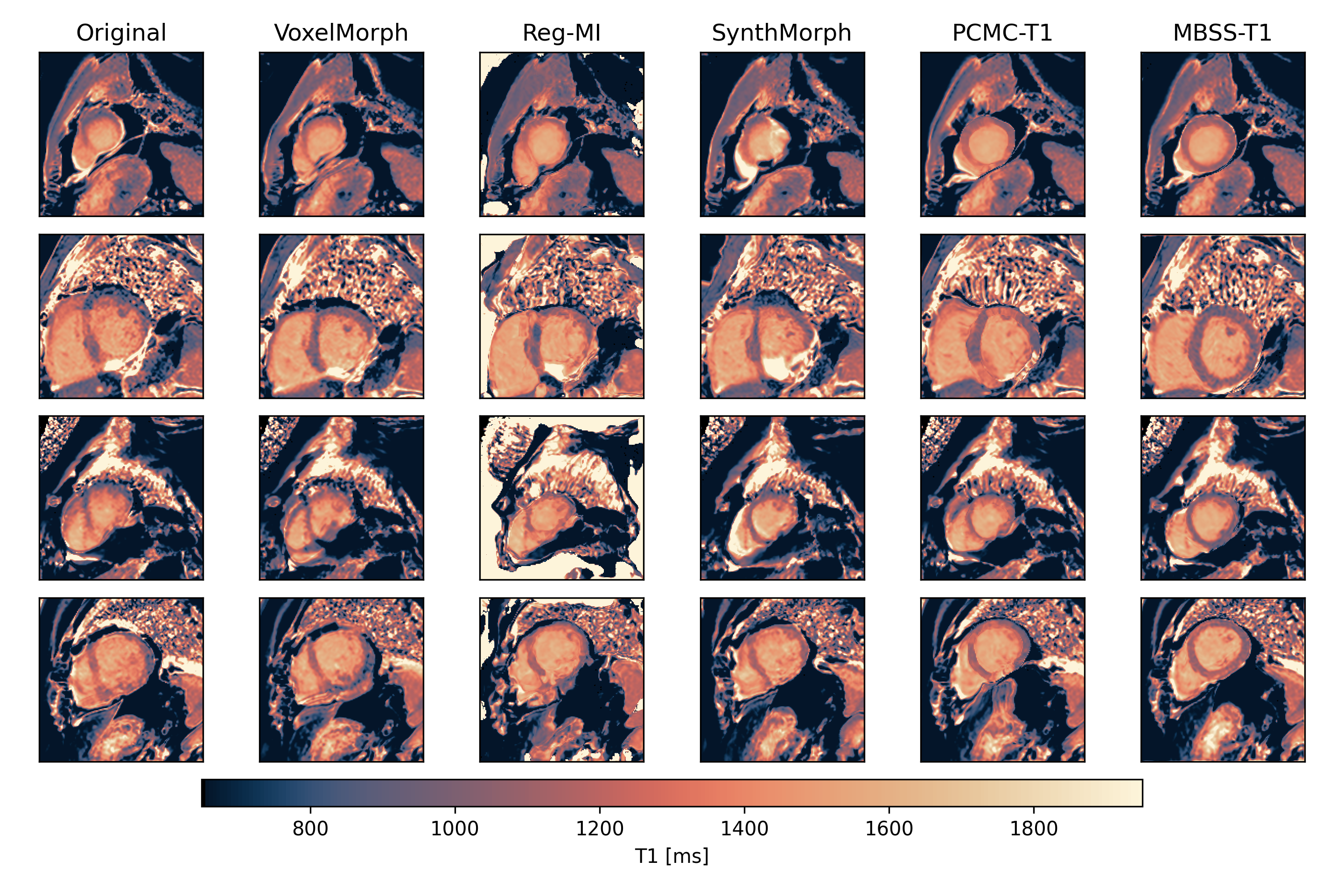}
\centering
\caption{Representative T1 maps computed with the different approaches. Our approach (MBSS-T1$_\textrm{STONE}$) demonstrates a clearer delineation between the blood and the muscle with a reduced partial volume effect, resulting in a more homogeneous mapping of the myocardium. The maps are presented in the colormap recommended by \cite{fuderer2025color}.}
\label{fig:T1_maps_stone}
\end{figure*}

\begin{figure*}[t!]\includegraphics[width=0.75\textwidth]{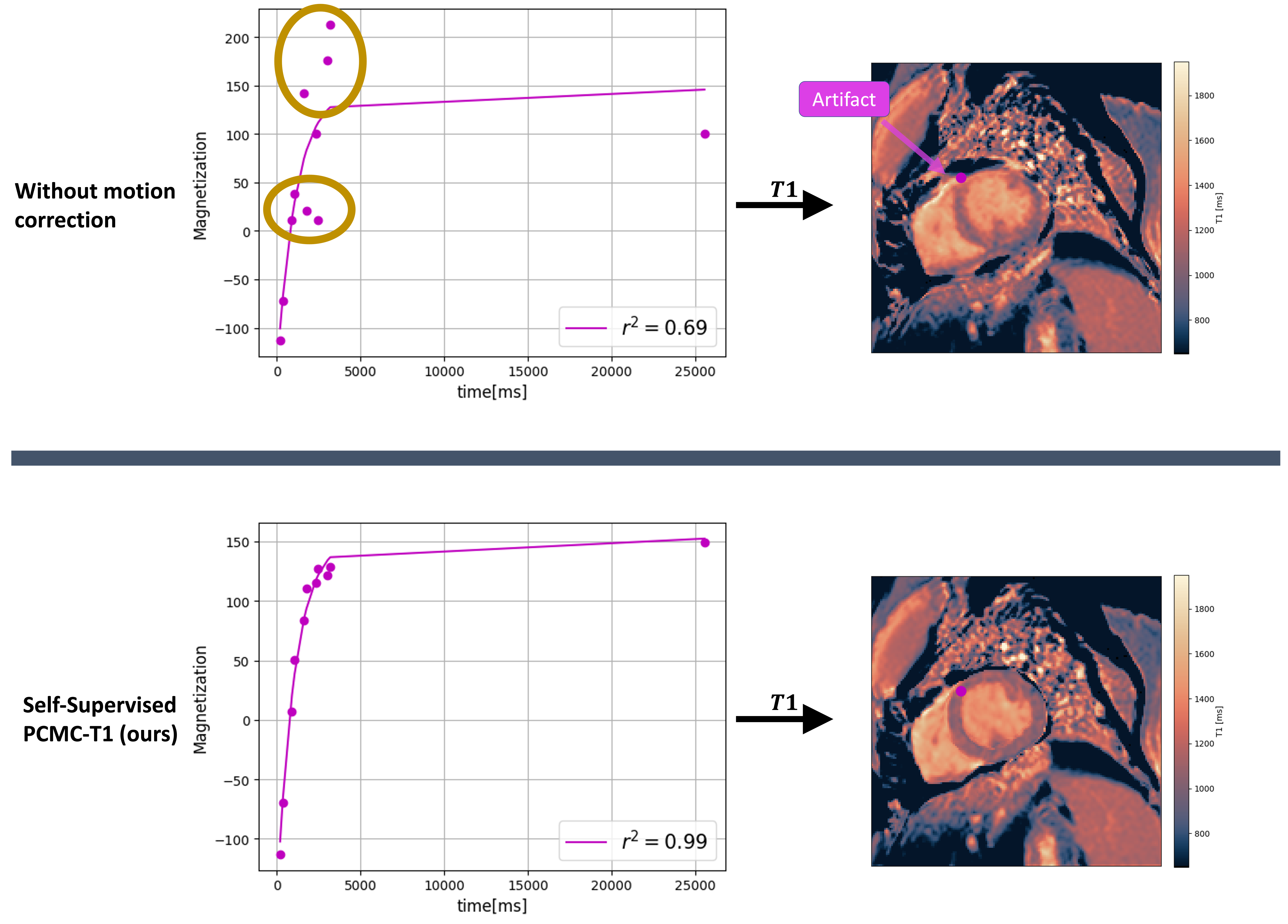}
\centering
\caption{Comparison of cardiac T1 mapping results on a STONE sequence with and without motion correction presented in the colormap recommended by \cite{fuderer2025color}. Without motion correction (top row), the T1 fit shows significant deviations in magnetization values (circled), leading to artifacts in the resulting T1 map. This is reflected in the lower \(R^2 = 0.69\). Using the MBSS-T1 method (bottom row), the T1 fit closely follows the expected curve, resulting in a higher \(R^2 = 0.99\) and a more accurate T1 map, with reduced artifacts in the corresponding region. This demonstrates the effectiveness of motion correction in improving the accuracy and quality of cardiac T1 mapping.}
\label{fig:fitting_comparison}
\end{figure*}

\noindent{\bf Clinical impact:}
Figure~\ref{fig:T1_maps_stone} presents several representative cases. The maps generated by the MBSS-T1$_\textrm{STONE}$ method demonstrate the highest quality, as evidenced by superior metrics. The rightmost column of Table~\ref{table:results_stone} summarizes the clinical impact assessment results for the MBSS-T1$_\textrm{STONE}$ method, which achieved the highest quality score of \(4.33 \pm 0.54\). This score reflects an improvement of approximately 10\% over the following best method, PCMC-T1, which had a score of \(3.93 \pm 0.78\). The difference in radiologist grading was statistically significant (p $ \ll 10^{-5}$). 
Furthermore, as demonstrated in Figure~\ref{fig:fitting_comparison}, the MBSS-T1$_\textrm{STONE}$ method achieves a closer fit to the expected magnetization curve (with \(R^2 = 0.99\)) and reduces artifacts, which may contribute to improved diagnostic confidence. These results suggest the MBSS-T1$_\textrm{STONE}$ method's potential utility in clinical settings where motion artifacts can impact the accuracy of cardiac T1 mapping.

\subsection{Exp. 2: Motion Correction in Free-Breathing and Breath-Hold MRI Using MOLLI}

{\bf Quantitative evaluation:} Table~\ref{table:results_MOLLI} summarizes our results for Exp 2 for both breath-hold (BH) and free-breathing (FB) scans. The MBSS-T1$_\textrm{MOLLI}$ method outperformed Synthmorph in terms of \(R^2\), Dice score, and Hausdorff distance.  Specifically, MBSS-T1$_\textrm{MOLLI}$ achieved an \(R^2\) value of \(0.994 \pm 0.017\) for BH and \(0.987 \pm 0.022\) for FB, representing improvements of approximately 0.1\% and 0.5\%, respectively, compared to Synthmorph. Additionally, MBSS-T1$_\textrm{MOLLI}$ produced a Dice score of \(0.954 \pm 0.083\) for BH and \(0.963 \pm 0.029\) for FB, reflecting improvements of approximately 3\% and 4.7\%, respectively, compared to Synthmorph. Furthermore, MBSS-T1$_\textrm{MOLLI}$ reduced the Hausdorff distance to \(2.64 \pm 4.77\) mm for BH and \(2.33 \pm 0.96\) mm for FB, yielding improvements of 29\% and 37\%, respectively, compared to Synthmorph.

\noindent{\bf Clinical impact:}
The rightmost column in both the breath-hold (BH) and free-breathing (FB) sections of Table~\ref{table:results_MOLLI} summarizes the clinical assessment results for the MOLLI sequence experiment. The MBSS-T1$_\textrm{MOLLI}$ method enhances the quality of T1 maps by effectively reducing motion artifacts, as reflected by the highest clinical scores in both BH and FB scans. Specifically, MBSS-T1$_\textrm{MOLLI}$ achieved mean scores of \(3.79 \pm 0.56\) for BH and \(4.1 \pm 0.61\) for FB scans. These scores represent improvements of approximately 20\% in BH and 17\% in FB compared to the next best method, Synthmorph, which had scores of \(3.15 \pm 0.491\) and \(3.5 \pm 0.58\), respectively.

\begin{table*}[t]
\centering
\caption{Comparison of motion correction methods for MOLLI Hold Breath and Free Breathing scenarios. Each method is evaluated using $R^2$, Dice, Hausdorff Distance (HD), and Clinical Score.}
\footnotesize
\begin{adjustbox}{width=\textwidth}
\begin{tabular}{|c|c|c|c|c|c|c|c|c|}
\hline
Method & \multicolumn{4}{c|}{Breath Hold} & \multicolumn{4}{c|}{Free Breathing} \\
\cline{2-9}
 & $R^2 \uparrow$ & Dice $\uparrow$ & HD [mm] $\downarrow$ & Clinical Score $\uparrow$ & $R^2 \uparrow$ & Dice $\uparrow$ & HD [mm] $\downarrow$ & Clinical Score $\uparrow$ \\
\hline
w.o motion correction & 0.991 $\pm$ 0.026 & 0.871 $\pm$ 0.131 & 5.21 $\pm$ 5.33 & 2.89 $\pm$ 0.89 & 
0.965 $\pm$ 0.054 & 0.851 $\pm$ 0.141 & 5.45 $\pm$ 2.36 & 2.81 $\pm$ 0.93 \\
Siemens MyoMap & N/A & N/A & N/A & 2.84 $\pm$ 0.66 & 
N/A & N/A & N/A & 3.28 $\pm$ 0.73 \\
SynthMorph & 0.993 $\pm$ 0.020 & 0.924 $\pm$ 0.087 & 3.75 $\pm$ 4.91 & 3.15 $\pm$ 0.62 & 
0.982 $\pm$ 0.026 & 0.919 $\pm$ 0.080 & 3.71 $\pm$ 1.51 & 3.5 $\pm$ 0.58 \\
MBSS-T1$_\textrm{MOLLI}$ & \textbf{0.994 $\pm$ 0.017} & \textbf{0.954 $\pm$ 0.083} & \textbf{2.64 $\pm$ 4.77} & \textbf{3.79 $\pm$ 0.56} & 
\textbf{0.987 $\pm$ 0.022} & \textbf{0.963 $\pm$ 0.029} & \textbf{2.33 $\pm$ 0.96} & \textbf{4.1 $\pm$ 0.61} \\
\hline
\end{tabular}
\end{adjustbox}
\label{table:results_MOLLI}
\end{table*}

\begin{figure}[t!]\includegraphics[width=0.46\textwidth]{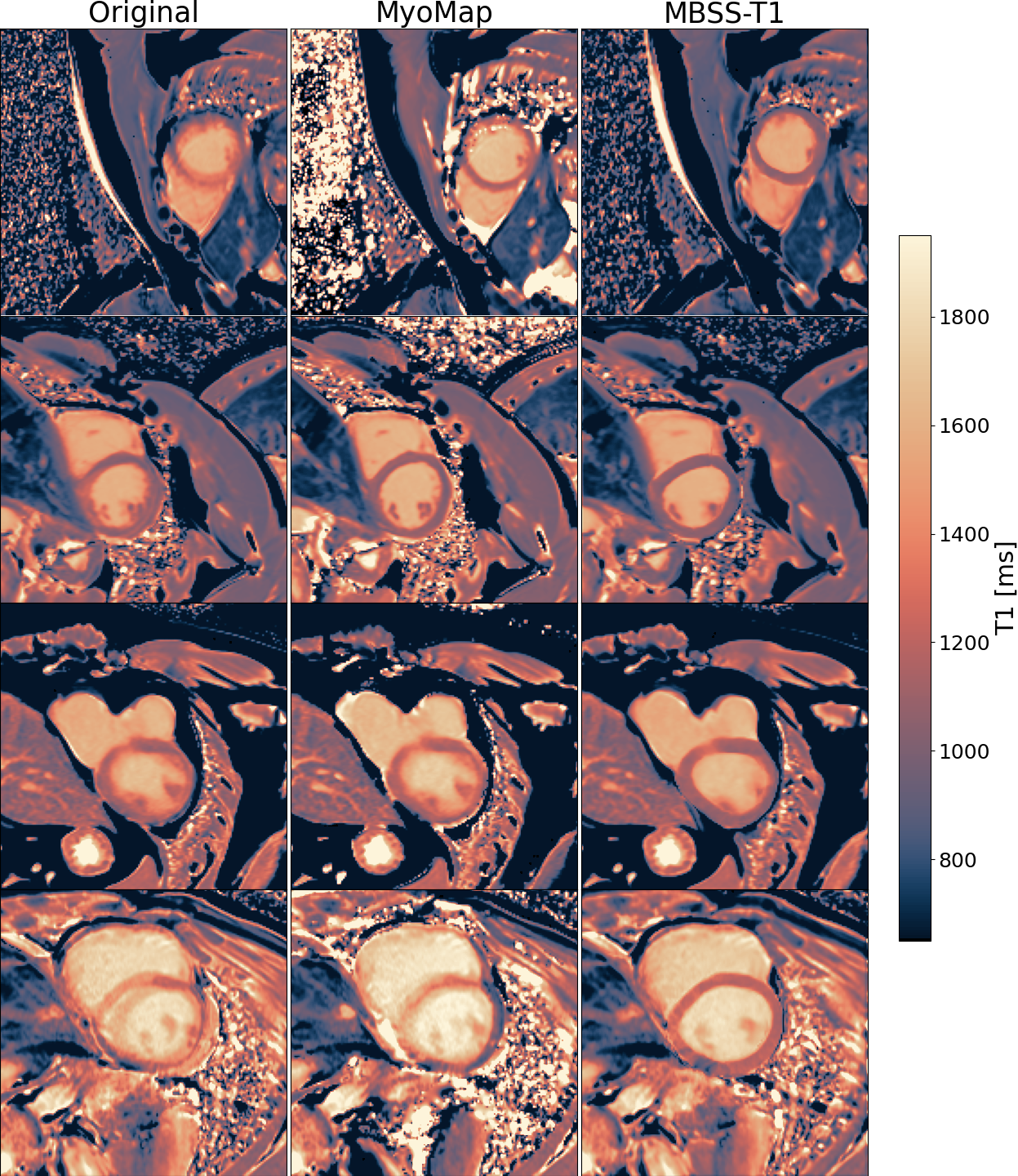}
\centering
\caption{Comparison of cardiac T1 mapping results for four patients (rows) using three motion correction methods: Original (left), Siemens MyoMap (middle), and MBSS-T1$_\textrm{MOLLI}$ (right). The maps are presented in the colormap recommended by \cite{fuderer2025color}. The top two rows represent T1 maps acquired using the MOLLI protocol during free breathing, while the bottom two rows represent T1 maps acquired using the MOLLI protocol during breath-hold. MBSS-T1$_\textrm{MOLLI}$ demonstrates superior motion correction, producing clearer T1 maps with fewer artifacts and better differentiation between the blood and myocardium.}
\label{fig:T1_maps_molli}
\end{figure}

Figure~\ref{fig:T1_maps_molli} presents the reduction in motion-related distortions achieved by the MBSS-T1$_\textrm{MOLLI}$ method. This improved motion correction results in more precise and reliable T1 maps with fewer artifacts.

\begin{figure}[t!]\includegraphics[width=\columnwidth]{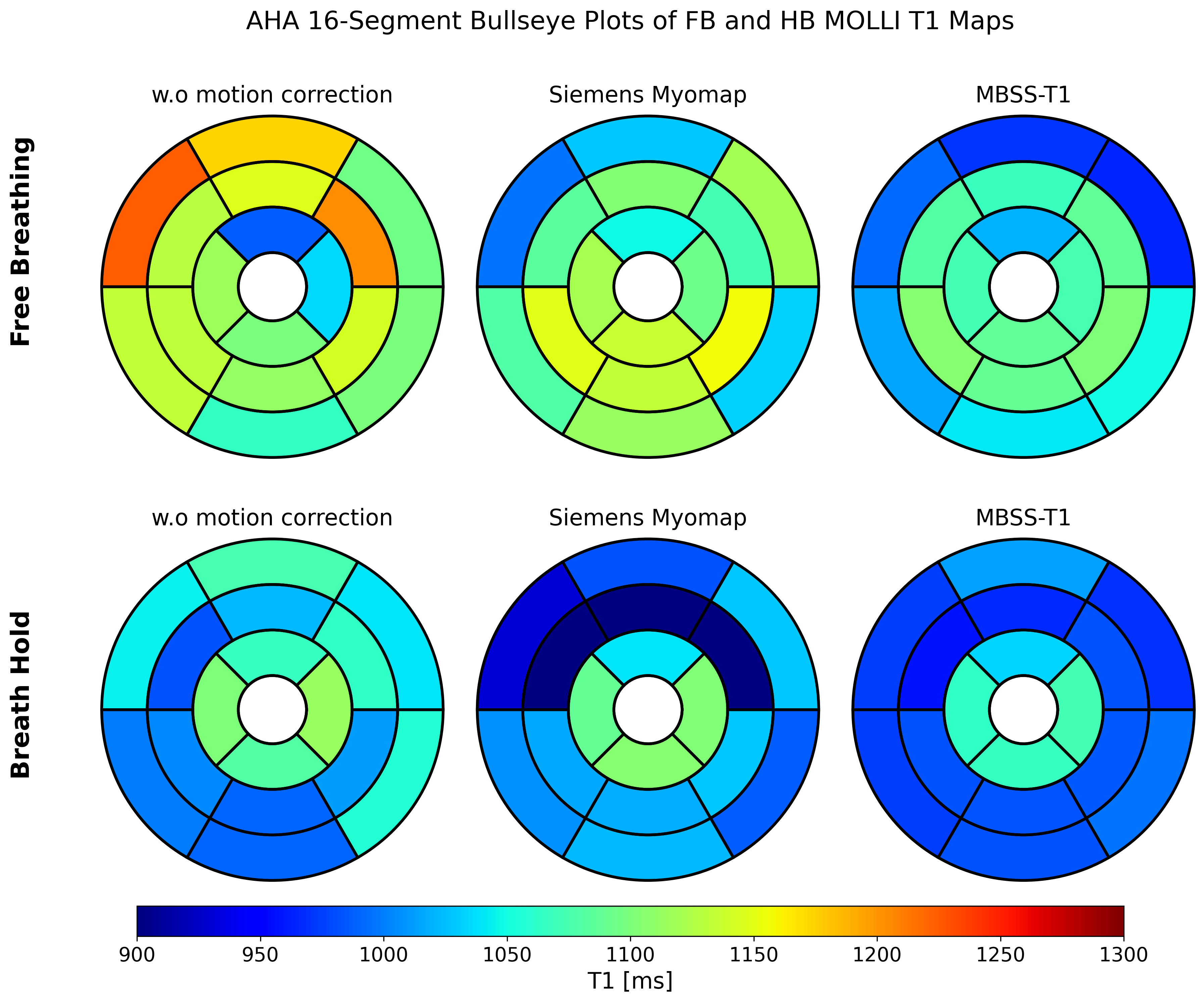}
\centering
\caption{Representative AHA 16-segment bullseye plots of myocardial T1 values for different motion correction methods and acquisition conditions. For both free-breathing and breath-hold acquisitions, the MBSS-T1$_\textrm{MOLLI}$ method generated more uniform T1 maps compared to the other methods.}
\label{fig:AHA_16_Bullseye_T1}
\end{figure}

Figure~\ref{fig:AHA_16_Bullseye_T1} shows the T1 values calculated by the different methods, presented using a 16-segment AHA bullseye plot for a representative subject. For both free-breathing and breath-hold acquisitions, our MBSS-T1$_\textrm{MOLLI}$ method produced more homogeneous maps compared to the other methods.

\subsection{Exp. 3: Consistency of cardiac T1 Mapping with Motion Correction in Test/ReTest MOLLI Dataset}

\begin{figure}[t!]\includegraphics[width=\columnwidth]{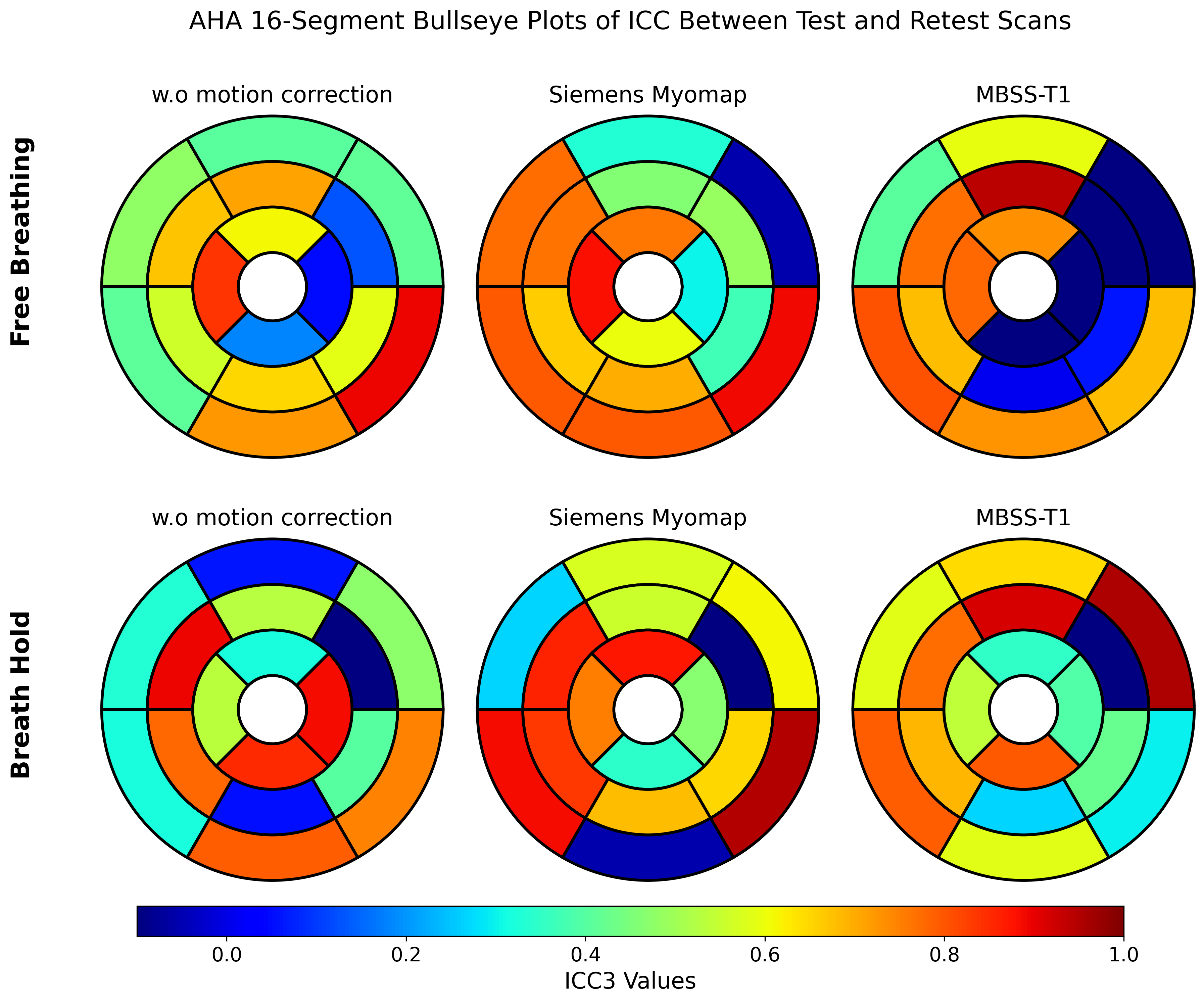}
\centering
\caption{AHA 16-segment bullseye plots of intraclass correlation coefficient (ICC) between test and retest scans for MOLLI cardiac T1 mapping. The plots compare ICC values for different motion correction techniques (without motion correction, Siemens MyoMap, and MBSS-T1) under free breathing and breath-hold conditions.}
\label{fig:AHA_16_Bullseye_ICC}
\end{figure}

Figure~\ref{fig:AHA_16_Bullseye_ICC} presents the ICC3 metrics for the mean myocardial T1 values across three motion correction approaches: no motion correction, Siemens MyoMap, and MBSS-T1$\textrm{MOLLI}$. Both motion correction methods, Siemens MyoMap and MBSS-T1$\textrm{MOLLI}$, demonstrated higher ICC values overall, indicating improved repeatability of the T1 maps. However, at the segmental level, neither method had a consistent advantage. In some segments, MBSS-T1$_\textrm{MOLLI}$ showed higher ICC values, while Siemens MyoMap performed better in others. Interestingly, for the free-breathing (FB) data, the motion correction algorithms yielded lower reproducibility than maps computed without motion correction for certain segments.

\section{Discussion}
This study introduced MBSS-T1, a subject-specific, self-supervised deep-learning approach for motion correction in cardiac T1 mapping. By integrating physical constraints through a signal model and anatomical constraints via segmentation-guided registration, the method addresses the challenges of motion artifacts in both free-breathing and breath-hold acquisitions. Unlike prior approaches requiring extensive training datasets, MBSS-T1 leverages a self-supervised framework, making it adaptable to diverse imaging protocols without requiring protocol-specific annotations.

The results demonstrated the effectiveness of MBSS-T1 across multiple datasets and acquisition sequences. In FB STONE acquisitions, MBSS-T1 outperformed baseline methods, achieving higher R$^2$ values, Dice scores, and improved Hausdorff distances. For MOLLI acquisitions, the approach generated more homogeneous T1 maps, with higher clinical scores for both BH and FB acquisitions, surpassing the performance of Siemens MyoMap and baseline method without motion correction. These results highlight MBSS-T1's robustness in mitigating motion artifacts and its potential to deliver clinically reliable T1 maps under challenging conditions.

The superior performance of MBSS-T1 compared to its preliminary version PCMC-T1, as reflected in both clinical score and segmentation metrics, can be attributed to two key factors. First, MBSS-T1 incorporates a pre-trained network specifically designed for segmentation tasks, enabling it to utilize segmentation results directly within its pipeline. This approach eliminates the need to simultaneously learn myocardium segmentation and registration, allowing the network to focus solely on the registration process and optimize the Dice loss more effectively. Second, MBSS-T1 employs a case-specific inference approach, solving the optimization problem individually for each test case rather than relying on a standard inference procedure, as is done in PCMC-T1. This tailored optimization strategy enhances robustness and accuracy, contributing to the significantly better performance of MBSS-T1. 

Interestingly, the FB scans with motion-corrected data achieved superior clinical scores compared to the BH scans, which is a counterintuitive result. The effect of the breath-hold can be observed in the higher clinical scores assigned to the fitting process without motion correction, reflecting the inherent quality of the acquired data.  We hypothesize that lower scores for the maps computed from BH scans using motion-correction algorithms may stem from the behavior of the motion-correction algorithms (Siemens MyoMap or MBSS-T1). For the BH scans, these algorithms likely detected minimal motion due to the breath hold, resulting in a limited capacity to correct any residual artifacts. As a result, minor artifacts may have persisted in the data. Conversely, the FB scans exhibited more substantial motion, which allowed the motion correction algorithms to address and mitigate a greater range of motion-related artifacts, ultimately improving the clinical scores.

It is important to note that the proposed method does not strictly guarantee anatomically correct deformations or a registration that fully compensates for underlying physiological motion. Instead, the method aims to align images to a reference state using physical and anatomical constraints that encourage expected signal decay behavior and foster realistic deformations. These constraints work to mitigate fitting errors that may arise due to motion, but their impact extends beyond motion-related artifacts to other inconsistencies in the data. Consequently, while the registration aligns the images in a manner suitable for the mapping network, it does not exclusively address physiological motion. 

Another important consideration in this work is the ability of the methods to generate diffeomorphic deformation fields, which ensure smooth and invertible transformations. While achieving full diffeomorphism is not explicitly required in our approach, the registration backbone based on the diffeomorphic Voxelmorph framework encourages but does not guarantee diffeomorphic flows, especially in the context of cardiac imaging with complex respiratory and cardiac motion. Notably, strict diffeomorphic properties can be achieved by replacing the registration backbone in our system with any other diffeomorphic registration framework. This adaptability allows the method to be tailored for applications where diffeomorphic deformations are a critical requirement.

Further, our method is designed to operate on magnitude images and focuses specifically on aligning single-shot images acquired at different time points, addressing inter-frame motion. However, it does not account for motion occurring during the acquisition of the single-shot images (intra-frame motion). Such motion can introduce distortions that are beyond the scope of our current approach. Future work could explore the integration of motion-resilient acquisition techniques or advanced intra-frame motion modeling to overcome this limitation and further enhance the robustness of cardiac T1 mapping.

The subject-specific nature of our approach indicates that the neural network employs a Memorization Learning strategy, focusing on specific details or particular features of the individual examples rather than adopting the common Pattern Learning strategy, where deep neural networks (DNNs) learn generalized features or common patterns \cite{wei2024memorization}. Future research is needed to investigate the generalization capacity of our approach, particularly when applied to larger and more diverse datasets.

While the proposed MBSS-T1 method demonstrates significant advancements in motion correction for cardiac T1 mapping, several limitations must be acknowledged to contextualize its findings and guide future improvements. First, the hyperparameters and the pre-trained segmentation network used in this study were specifically optimized for the free-breathing STONE sequence and subsequently evaluated on both the STONE free-breathing and MOLLI breath-hold and free-breathing sequences. This approach was necessitated by the limited number of MOLLI cases available in our dataset, which constrained our ability to perform robust optimization tailored specifically to MOLLI data. As a result, while our method demonstrated promising results on the MOLLI dataset, it is likely that these results could be further improved through hyperparameter optimization specifically designed for MOLLI data. 

In addition, while our study demonstrates the potential of MBSS-T1 for motion-robust cardiac T1 mapping, particularly under FB conditions and suboptimal compliance with BH instructions, we acknowledge that our approach has limitations in addressing sub-optimal echo-triggering due to arrhythmias. Specifically, the assumed signal model may fail in the presence of arrhythmias due to spin history effects~\cite{fitts2013arrhythmia}.

Further, the automated selection of the reference image based on segmentation performance may introduce registration bias, favoring the anatomical features of the chosen image. While this enhances segmentation accuracy, it may not represent the average anatomical state. Future work could explore alternatives, such as using an average template or dynamic reference selection, to reduce this potential bias.

Finally, there is room for further optimization of the network architecture to enhance its performance and generalization. One promising area for future exploration is the incorporation of different normalization layers, which have the potential to stabilize training, improve convergence, and enhance the network's robustness.

\section{Conclusion}
In this study, we proposed MBSS-T1, a subject-specific, self-supervised motion correction method for robust cardiac T1 mapping. By integrating physical constraints through a signal model and anatomical constraints via segmentation, MBSS-T1 effectively mitigates motion artifacts in both free-breathing and breath-hold acquisitions. The method demonstrated superior performance compared to baseline approaches, producing more homogeneous T1 maps and achieving higher clinical scores and segmentation metrics. Importantly, the self-supervised nature of MBSS-T1 enables its adaptability to diverse imaging protocols without requiring large annotated datasets. Future work will focus on addressing current limitations, including exploring advanced motion modeling and optimization strategies, to further enhance the method's robustness and clinical applicability.

\section*{Acknowledgments}
This work was supported in part by research grants from the Israel-US Binational Science Foundation, the Israeli Ministry of Science and Technology, the Israel Innovation Authority, and the joint Microsoft Education and the Israel Inter-university Computation Center (IUCC) program.

\section*{Declaration of Generative AI}
During the preparation of this work, the author(s) used ChatGPT in order to improve readability. After using this tool/service, the author(s) reviewed and edited the content as needed and take(s) full responsibility for the content of the publication.

\bibliographystyle{model2-names.bst}\biboptions{authoryear}
\bibliography{refs}

\begin{thebibliography}{41}
\expandafter\ifx\csname natexlab\endcsname\relax\def\natexlab#1{#1}\fi
\providecommand{\url}[1]{\texttt{#1}}
\providecommand{\href}[2]{#2}
\providecommand{\path}[1]{#1}
\providecommand{\DOIprefix}{doi:}
\providecommand{\ArXivprefix}{arXiv:}
\providecommand{\URLprefix}{URL: }
\providecommand{\Pubmedprefix}{pmid:}
\providecommand{\doi}[1]{\href{http://dx.doi.org/#1}{\path{#1}}}
\providecommand{\Pubmed}[1]{\href{pmid:#1}{\path{#1}}}
\providecommand{\bibinfo}[2]{#2}
\ifx\xfnm\relax \def\xfnm[#1]{\unskip,\space#1}\fi
\bibitem[{Arava et~al.(2021)Arava, Masarwy, Khawaled and Freiman}]{arava2021deep}
\bibinfo{author}{Arava, D.}, \bibinfo{author}{Masarwy, M.}, \bibinfo{author}{Khawaled, S.}, \bibinfo{author}{Freiman, M.}, \bibinfo{year}{2021}.
\newblock \bibinfo{title}{Deep-learning based motion correction for myocardial {T1} mapping}, in: \bibinfo{booktitle}{2021 IEEE International Conference on Microwaves, Antennas, Communications and Electronic Systems (COMCAS)}, \bibinfo{organization}{IEEE}. pp. \bibinfo{pages}{55--59}.
\bibitem[{Balakrishnan et~al.(2019)Balakrishnan, Zhao, Sabuncu, Guttag and Dalca}]{balakrishnan2019voxelmorph}
\bibinfo{author}{Balakrishnan, G.}, \bibinfo{author}{Zhao, A.}, \bibinfo{author}{Sabuncu, M.R.}, \bibinfo{author}{Guttag, J.}, \bibinfo{author}{Dalca, A.V.}, \bibinfo{year}{2019}.
\newblock \bibinfo{title}{Voxelmorph: a learning framework for deformable medical image registration}.
\newblock \bibinfo{journal}{IEEE Transactions on medical imaging} \bibinfo{volume}{38}, \bibinfo{pages}{1788--1800}.
\bibitem[{Dalca et~al.(2019)Dalca, Balakrishnan, Guttag and Sabuncu}]{dalca2019unsupervised}
\bibinfo{author}{Dalca, A.V.}, \bibinfo{author}{Balakrishnan, G.}, \bibinfo{author}{Guttag, J.}, \bibinfo{author}{Sabuncu, M.R.}, \bibinfo{year}{2019}.
\newblock \bibinfo{title}{Unsupervised learning of probabilistic diffeomorphic registration for images and surfaces}.
\newblock \bibinfo{journal}{Medical image analysis} \bibinfo{volume}{57}, \bibinfo{pages}{226--236}.
\bibitem[{El-Rewaidy et~al.(2018)El-Rewaidy, Nezafat, Jang, Nakamori, Fahmy and Nezafat}]{el2018nonrigid}
\bibinfo{author}{El-Rewaidy, H.}, \bibinfo{author}{Nezafat, M.}, \bibinfo{author}{Jang, J.}, \bibinfo{author}{Nakamori, S.}, \bibinfo{author}{Fahmy, A.S.}, \bibinfo{author}{Nezafat, R.}, \bibinfo{year}{2018}.
\newblock \bibinfo{title}{Nonrigid active shape model-based registration framework for motion correction of cardiac {T1} mapping}.
\newblock \bibinfo{journal}{Magnetic resonance in medicine} \bibinfo{volume}{80}, \bibinfo{pages}{780--791}.
\bibitem[{Fahmy(2019)}]{DHEUAV_2019}
\bibinfo{author}{Fahmy, A.S.}, \bibinfo{year}{2019}.
\newblock \bibinfo{title}{Replication data for: Automated analysis of cardiovascular magnetic resonance myocardial native {T1} mapping images using fully convolutional neural networks}.
\newblock \URLprefix \url{https://doi.org/10.7910/DVN/DHEUAV}, \DOIprefix\doi{10.7910/DVN/DHEUAV}.
\bibitem[{Fitts et~al.(2013)Fitts, Breton, Kholmovski, Dosdall, Vijayakumar, Hong, Ranjan, Marrouche, Axel and Kim}]{fitts2013arrhythmia}
\bibinfo{author}{Fitts, M.}, \bibinfo{author}{Breton, E.}, \bibinfo{author}{Kholmovski, E.G.}, \bibinfo{author}{Dosdall, D.J.}, \bibinfo{author}{Vijayakumar, S.}, \bibinfo{author}{Hong, K.P.}, \bibinfo{author}{Ranjan, R.}, \bibinfo{author}{Marrouche, N.F.}, \bibinfo{author}{Axel, L.}, \bibinfo{author}{Kim, D.}, \bibinfo{year}{2013}.
\newblock \bibinfo{title}{Arrhythmia insensitive rapid cardiac t1 mapping pulse sequence}.
\newblock \bibinfo{journal}{Magnetic resonance in medicine} \bibinfo{volume}{70}, \bibinfo{pages}{1274--1282}.
\bibitem[{Fuderer et~al.(2025)Fuderer, Wichtmann, Crameri, de~Souza, Bae{\ss}ler, Gulani, Wang, Poot, de~Boer, Cashmore et~al.}]{fuderer2025color}
\bibinfo{author}{Fuderer, M.}, \bibinfo{author}{Wichtmann, B.}, \bibinfo{author}{Crameri, F.}, \bibinfo{author}{de~Souza, N.M.}, \bibinfo{author}{Bae{\ss}ler, B.}, \bibinfo{author}{Gulani, V.}, \bibinfo{author}{Wang, M.}, \bibinfo{author}{Poot, D.}, \bibinfo{author}{de~Boer, R.}, \bibinfo{author}{Cashmore, M.}, et~al., \bibinfo{year}{2025}.
\newblock \bibinfo{title}{Color-map recommendation for mr relaxometry maps}.
\newblock \bibinfo{journal}{Magnetic Resonance in Medicine} \bibinfo{volume}{93}, \bibinfo{pages}{490--506}.
\bibitem[{van~de Giessen et~al.(2013)van~de Giessen, Tao, van~der Geest and Lelieveldt}]{van2013model}
\bibinfo{author}{van~de Giessen, M.}, \bibinfo{author}{Tao, Q.}, \bibinfo{author}{van~der Geest, R.J.}, \bibinfo{author}{Lelieveldt, B.P.}, \bibinfo{year}{2013}.
\newblock \bibinfo{title}{Model-based alignment of look-locker {MRI} sequences for calibrated myocardical scar tissue quantification}, in: \bibinfo{booktitle}{2013 IEEE 10th International Symposium on Biomedical Imaging (ISBI)}, pp. \bibinfo{pages}{1038--1041}.
\bibitem[{Gonzales et~al.(2022)Gonzales, Zhang, Papie{\.z}, Werys, Lukaschuk, Popescu, Burrage, Shanmuganathan, Ferreira and Piechnik}]{gonzales2022fast}
\bibinfo{author}{Gonzales, R.A.}, \bibinfo{author}{Zhang, Q.}, \bibinfo{author}{Papie{\.z}, B.}, \bibinfo{author}{Werys, K.}, \bibinfo{author}{Lukaschuk, E.}, \bibinfo{author}{Popescu, I.A.}, \bibinfo{author}{Burrage, M.K.}, \bibinfo{author}{Shanmuganathan, M.}, \bibinfo{author}{Ferreira, V.M.}, \bibinfo{author}{Piechnik, S.K.}, \bibinfo{year}{2022}.
\newblock \bibinfo{title}{Fast and robust motion correction of cardiovascular magnetic resonance {T1}-mapping using data-driven convolutional neural networks for generalisability}, in: \bibinfo{booktitle}{Proc. Society for Cardiovascular Magnetic Resonance Virtual Annual Scientific Sessions (SCMR)}, \bibinfo{publisher}{Society for Cardiovascular Magnetic Resonance}.
\bibitem[{Gonzales et~al.(2021)Gonzales, Zhang, Papie{\.z}, Werys, Lukaschuk, Popescu, Burrage, Shanmuganathan, Ferreira and Piechnik}]{gonzales2021moconet}
\bibinfo{author}{Gonzales, R.A.}, \bibinfo{author}{Zhang, Q.}, \bibinfo{author}{Papie{\.z}, B.W.}, \bibinfo{author}{Werys, K.}, \bibinfo{author}{Lukaschuk, E.}, \bibinfo{author}{Popescu, I.A.}, \bibinfo{author}{Burrage, M.K.}, \bibinfo{author}{Shanmuganathan, M.}, \bibinfo{author}{Ferreira, V.M.}, \bibinfo{author}{Piechnik, S.K.}, \bibinfo{year}{2021}.
\newblock \bibinfo{title}{{MOCOnet}: robust motion correction of cardiovascular magnetic resonance {T1} mapping using convolutional neural networks}.
\newblock \bibinfo{journal}{Frontiers in Cardiovascular Medicine} , \bibinfo{pages}{1689}.
\bibitem[{Guo et~al.(2022)Guo, El-Rewaidy, Assana, Cai, Amyar, Chow, Bi, Yankama, Cirillo, Pierce et~al.}]{guo2022accelerated}
\bibinfo{author}{Guo, R.}, \bibinfo{author}{El-Rewaidy, H.}, \bibinfo{author}{Assana, S.}, \bibinfo{author}{Cai, X.}, \bibinfo{author}{Amyar, A.}, \bibinfo{author}{Chow, K.}, \bibinfo{author}{Bi, X.}, \bibinfo{author}{Yankama, T.}, \bibinfo{author}{Cirillo, J.}, \bibinfo{author}{Pierce, P.}, et~al., \bibinfo{year}{2022}.
\newblock \bibinfo{title}{Accelerated cardiac {T1} mapping in four heartbeats with inline {MyoMapNet}: a deep learning-based {T1} estimation approach}.
\newblock \bibinfo{journal}{Journal of Cardiovascular Magnetic Resonance} \bibinfo{volume}{24}, \bibinfo{pages}{1--15}.
\bibitem[{Guyader et~al.(2016)Guyader, Huizinga, Fortunati, Poot, Van~Kranenburg, Veenland, Paulides, Niessen and Klein}]{guyader2016total}
\bibinfo{author}{Guyader, J.M.}, \bibinfo{author}{Huizinga, W.}, \bibinfo{author}{Fortunati, V.}, \bibinfo{author}{Poot, D.H.}, \bibinfo{author}{Van~Kranenburg, M.}, \bibinfo{author}{Veenland, J.F.}, \bibinfo{author}{Paulides, M.M.}, \bibinfo{author}{Niessen, W.J.}, \bibinfo{author}{Klein, S.}, \bibinfo{year}{2016}.
\newblock \bibinfo{title}{Total correlation-based groupwise image registration for quantitative {MRI}}, in: \bibinfo{booktitle}{Proc. IEEE conference on computer vision and pattern recognition workshops}, pp. \bibinfo{pages}{186--193}.
\bibitem[{Hanania et~al.(2022)Hanania, Barkat, Cohen, Azhari and Freiman}]{hananiastylereg}
\bibinfo{author}{Hanania, E.}, \bibinfo{author}{Barkat, L.}, \bibinfo{author}{Cohen, I.}, \bibinfo{author}{Azhari, H.}, \bibinfo{author}{Freiman, M.}, \bibinfo{year}{2022}.
\newblock \bibinfo{title}{{StyleReg}: Style transfer as a preprocessing step for myocardial {T1} mapping}, in: \bibinfo{booktitle}{Medical Imaging meets NeurIPS (Med-NeurIPS-2022) Workshop at the 36th Conference on Neural Information Processing Systems}, \bibinfo{address}{New Orleans, USA}. pp. \bibinfo{pages}{1--5}.
\bibitem[{Hanania et~al.(2023a)Hanania, Barkat, Cohen, Azhari and Freiman}]{hanania2023groupT1}
\bibinfo{author}{Hanania, E.}, \bibinfo{author}{Barkat, L.}, \bibinfo{author}{Cohen, I.}, \bibinfo{author}{Azhari, H.}, \bibinfo{author}{Freiman, M.}, \bibinfo{year}{2023}a.
\newblock \bibinfo{title}{Deep-learning-based group-wise motion correction for myocardial {T1} mapping}, in: \bibinfo{booktitle}{Proc. ISMRM \& SMRT Annual Meeting \& Exhibition, Toronto, Canada}.
\bibitem[{Hanania et~al.(2023b)Hanania, Volovik, Barkat, Cohen and Freiman}]{hanania2023pcmc}
\bibinfo{author}{Hanania, E.}, \bibinfo{author}{Volovik, I.}, \bibinfo{author}{Barkat, L.}, \bibinfo{author}{Cohen, I.}, \bibinfo{author}{Freiman, M.}, \bibinfo{year}{2023}b.
\newblock \bibinfo{title}{{PCMC-T1}: Free-breathing myocardial {T1} mapping with physically-constrained motion correction}, in: \bibinfo{booktitle}{Proc. 2023 Int. Conf. Medical Image Computing and Computer-Aided Intervention (MICCAI 2023), Lecture Notes in Computer Science}, \bibinfo{organization}{Springer}. pp. \bibinfo{pages}{226--235}.
\bibitem[{Hering et~al.(2019)Hering, Kuckertz, Heldmann and Heinrich}]{hering2019enhancing}
\bibinfo{author}{Hering, A.}, \bibinfo{author}{Kuckertz, S.}, \bibinfo{author}{Heldmann, S.}, \bibinfo{author}{Heinrich, M.P.}, \bibinfo{year}{2019}.
\newblock \bibinfo{title}{Enhancing label-driven deep deformable image registration with local distance metrics for state-of-the-art cardiac motion tracking}, in: \bibinfo{editor}{Handels, H.}, \bibinfo{editor}{Deserno, T.M.}, \bibinfo{editor}{Maier, A.}, \bibinfo{editor}{Maier-Hein, K.H.}, \bibinfo{editor}{Palm, C.}, \bibinfo{editor}{Tolxdorff, T.} (Eds.), \bibinfo{booktitle}{Bildverarbeitung f{\"u}r die Medizin 2019}, pp. \bibinfo{pages}{309--314}.
\bibitem[{Hoffmann et~al.(2021)Hoffmann, Billot, Greve, Iglesias, Fischl and Dalca}]{hoffmann2021synthmorph}
\bibinfo{author}{Hoffmann, M.}, \bibinfo{author}{Billot, B.}, \bibinfo{author}{Greve, D.N.}, \bibinfo{author}{Iglesias, J.E.}, \bibinfo{author}{Fischl, B.}, \bibinfo{author}{Dalca, A.V.}, \bibinfo{year}{2021}.
\newblock \bibinfo{title}{Synthmorph: learning contrast-invariant registration without acquired images}.
\newblock \bibinfo{journal}{IEEE Transactions on medical imaging} \bibinfo{volume}{41}, \bibinfo{pages}{543--558}.
\bibitem[{Huizinga et~al.(2016)Huizinga, Poot, Guyader, Klaassen, Coolen, van Kranenburg, Van~Geuns, Uitterdijk, Polfliet, Vandemeulebroucke et~al.}]{huizinga2016pca}
\bibinfo{author}{Huizinga, W.}, \bibinfo{author}{Poot, D.H.}, \bibinfo{author}{Guyader, J.M.}, \bibinfo{author}{Klaassen, R.}, \bibinfo{author}{Coolen, B.F.}, \bibinfo{author}{van Kranenburg, M.}, \bibinfo{author}{Van~Geuns, R.}, \bibinfo{author}{Uitterdijk, A.}, \bibinfo{author}{Polfliet, M.}, \bibinfo{author}{Vandemeulebroucke, J.}, et~al., \bibinfo{year}{2016}.
\newblock \bibinfo{title}{{PCA}-based groupwise image registration for quantitative {MRI}}.
\newblock \bibinfo{journal}{Medical image analysis} \bibinfo{volume}{29}, \bibinfo{pages}{65--78}.
\bibitem[{Isensee et~al.(2021)Isensee, Jaeger, Kohl, Petersen and Maier-Hein}]{isensee2021nnu}
\bibinfo{author}{Isensee, F.}, \bibinfo{author}{Jaeger, P.F.}, \bibinfo{author}{Kohl, S.A.}, \bibinfo{author}{Petersen, J.}, \bibinfo{author}{Maier-Hein, K.H.}, \bibinfo{year}{2021}.
\newblock \bibinfo{title}{nnu-net: a self-configuring method for deep learning-based biomedical image segmentation}.
\newblock \bibinfo{journal}{Nature methods} \bibinfo{volume}{18}, \bibinfo{pages}{203--211}.
\bibitem[{K{\"u}stner et~al.(2021)K{\"u}stner, Pan, Qi, Cruz, Gilliam, Blu, Yang, Gatidis, Botnar and Prieto}]{kustner2021lapnet}
\bibinfo{author}{K{\"u}stner, T.}, \bibinfo{author}{Pan, J.}, \bibinfo{author}{Qi, H.}, \bibinfo{author}{Cruz, G.}, \bibinfo{author}{Gilliam, C.}, \bibinfo{author}{Blu, T.}, \bibinfo{author}{Yang, B.}, \bibinfo{author}{Gatidis, S.}, \bibinfo{author}{Botnar, R.}, \bibinfo{author}{Prieto, C.}, \bibinfo{year}{2021}.
\newblock \bibinfo{title}{Lapnet: Non-rigid registration derived in k-space for magnetic resonance imaging}.
\newblock \bibinfo{journal}{IEEE Transactions on medical imaging} \bibinfo{volume}{40}, \bibinfo{pages}{3686--3697}.
\bibitem[{Li et~al.(2021)Li, Wang, Qi, Hu, Chen, Yang, Qiao, Sun, Wang, Zhao et~al.}]{li2021deep}
\bibinfo{author}{Li, Y.}, \bibinfo{author}{Wang, Y.}, \bibinfo{author}{Qi, H.}, \bibinfo{author}{Hu, Z.}, \bibinfo{author}{Chen, Z.}, \bibinfo{author}{Yang, R.}, \bibinfo{author}{Qiao, H.}, \bibinfo{author}{Sun, J.}, \bibinfo{author}{Wang, T.}, \bibinfo{author}{Zhao, X.}, et~al., \bibinfo{year}{2021}.
\newblock \bibinfo{title}{Deep learning--enhanced {T1} mapping with spatial-temporal and physical constraint}.
\newblock \bibinfo{journal}{Magnetic Resonance in Medicine} \bibinfo{volume}{86}, \bibinfo{pages}{1647--1661}.
\bibitem[{Li et~al.(2022)Li, Wu, Qi, Si, Ding and Chen}]{li2022motion}
\bibinfo{author}{Li, Y.}, \bibinfo{author}{Wu, C.}, \bibinfo{author}{Qi, H.}, \bibinfo{author}{Si, D.}, \bibinfo{author}{Ding, H.}, \bibinfo{author}{Chen, H.}, \bibinfo{year}{2022}.
\newblock \bibinfo{title}{Motion correction for native myocardial {T1} mapping using self-supervised deep learning registration with contrast separation}.
\newblock \bibinfo{journal}{NMR in Biomedicine} \bibinfo{volume}{35}, \bibinfo{pages}{e4775}.
\bibitem[{McGraw and Wong(1996)}]{mcgraw1996forming}
\bibinfo{author}{McGraw, K.O.}, \bibinfo{author}{Wong, S.P.}, \bibinfo{year}{1996}.
\newblock \bibinfo{title}{Forming inferences about some intraclass correlation coefficients.}
\newblock \bibinfo{journal}{Psychological methods} \bibinfo{volume}{1}, \bibinfo{pages}{30}.
\bibitem[{Morales et~al.(2019)Morales, Izquierdo-Garcia, Aganj, Kalpathy-Cramer, Rosen and Catana}]{morales2019implementation}
\bibinfo{author}{Morales, M.A.}, \bibinfo{author}{Izquierdo-Garcia, D.}, \bibinfo{author}{Aganj, I.}, \bibinfo{author}{Kalpathy-Cramer, J.}, \bibinfo{author}{Rosen, B.R.}, \bibinfo{author}{Catana, C.}, \bibinfo{year}{2019}.
\newblock \bibinfo{title}{Implementation and validation of a three-dimensional cardiac motion estimation network}.
\newblock \bibinfo{journal}{Radiology: Artificial Intelligence} \bibinfo{volume}{1}, \bibinfo{pages}{e180080}.
\bibitem[{Pan et~al.(2024)Pan, Huang, Yu, Peng, Chuang, Lin, Chung and Wu}]{pan2024virtual}
\bibinfo{author}{Pan, N.Y.}, \bibinfo{author}{Huang, T.Y.}, \bibinfo{author}{Yu, J.J.}, \bibinfo{author}{Peng, H.H.}, \bibinfo{author}{Chuang, T.C.}, \bibinfo{author}{Lin, Y.R.}, \bibinfo{author}{Chung, H.W.}, \bibinfo{author}{Wu, M.T.}, \bibinfo{year}{2024}.
\newblock \bibinfo{title}{Virtual {MOLLI} target: Generative adversarial networks toward improved motion correction in {MRI} myocardial {T1} mapping}.
\newblock \bibinfo{journal}{Journal of Magnetic Resonance Imaging} .
\bibitem[{Qi et~al.(2021)Qi, Hajhosseiny, Cruz, Kuestner, Kunze, Neji, Botnar and Prieto}]{qi2021end}
\bibinfo{author}{Qi, H.}, \bibinfo{author}{Hajhosseiny, R.}, \bibinfo{author}{Cruz, G.}, \bibinfo{author}{Kuestner, T.}, \bibinfo{author}{Kunze, K.}, \bibinfo{author}{Neji, R.}, \bibinfo{author}{Botnar, R.}, \bibinfo{author}{Prieto, C.}, \bibinfo{year}{2021}.
\newblock \bibinfo{title}{End-to-end deep learning nonrigid motion-corrected reconstruction for highly accelerated free-breathing coronary mra}.
\newblock \bibinfo{journal}{Magnetic Resonance in Medicine} \bibinfo{volume}{86}, \bibinfo{pages}{1983--1996}.
\bibitem[{Qin et~al.(2018)Qin, Bai, Schlemper, Petersen, Piechnik, Neubauer and Rueckert}]{qin2018joint}
\bibinfo{author}{Qin, C.}, \bibinfo{author}{Bai, W.}, \bibinfo{author}{Schlemper, J.}, \bibinfo{author}{Petersen, S.E.}, \bibinfo{author}{Piechnik, S.K.}, \bibinfo{author}{Neubauer, S.}, \bibinfo{author}{Rueckert, D.}, \bibinfo{year}{2018}.
\newblock \bibinfo{title}{Joint learning of motion estimation and segmentation for cardiac {MR} image sequences}, in: \bibinfo{booktitle}{Proc. 2018 Int. Conf. Medical Image Computing and Computer-Aided Intervention (MICCAI 2018), Lecture Notes in Computer Science}, \bibinfo{organization}{Springer}. pp. \bibinfo{pages}{472--480}.
\bibitem[{Roujol et~al.(2015)Roujol, Foppa, Weing{\"a}rtner, Manning and Nezafat}]{roujol2015adaptive}
\bibinfo{author}{Roujol, S.}, \bibinfo{author}{Foppa, M.}, \bibinfo{author}{Weing{\"a}rtner, S.}, \bibinfo{author}{Manning, W.J.}, \bibinfo{author}{Nezafat, R.}, \bibinfo{year}{2015}.
\newblock \bibinfo{title}{Adaptive registration of varying contrast-weighted images for improved tissue characterization ({ARCTIC}): application to {T1} mapping}.
\newblock \bibinfo{journal}{Magnetic resonance in medicine} \bibinfo{volume}{73}, \bibinfo{pages}{1469--1482}.
\bibitem[{Roujol et~al.(2014)Roujol, Weing{\"a}rtner, Foppa, Chow, Kawaji, Ngo, Kellman, Manning, Thompson and Nezafat}]{roujol2014accuracy}
\bibinfo{author}{Roujol, S.}, \bibinfo{author}{Weing{\"a}rtner, S.}, \bibinfo{author}{Foppa, M.}, \bibinfo{author}{Chow, K.}, \bibinfo{author}{Kawaji, K.}, \bibinfo{author}{Ngo, L.H.}, \bibinfo{author}{Kellman, P.}, \bibinfo{author}{Manning, W.J.}, \bibinfo{author}{Thompson, R.B.}, \bibinfo{author}{Nezafat, R.}, \bibinfo{year}{2014}.
\newblock \bibinfo{title}{Accuracy, precision, and reproducibility of four {T1} mapping sequences: a head-to-head comparison of {MOLLI, ShMOLLI, SASHA}, and {SAPPHIRE}}.
\newblock \bibinfo{journal}{Radiology} \bibinfo{volume}{272}, \bibinfo{pages}{683--689}.
\bibitem[{Schelbert and Messroghli(2016)}]{schelbert2016state}
\bibinfo{author}{Schelbert, E.B.}, \bibinfo{author}{Messroghli, D.R.}, \bibinfo{year}{2016}.
\newblock \bibinfo{title}{State of the art: clinical applications of cardiac {T1} mapping}.
\newblock \bibinfo{journal}{Radiology} \bibinfo{volume}{278}, \bibinfo{pages}{658--676}.
\bibitem[{Tao et~al.(2018)Tao, van~der Tol, Berendsen, Paiman, Lamb and van~der Geest}]{tao2018robust}
\bibinfo{author}{Tao, Q.}, \bibinfo{author}{van~der Tol, P.}, \bibinfo{author}{Berendsen, F.F.}, \bibinfo{author}{Paiman, E.H.}, \bibinfo{author}{Lamb, H.J.}, \bibinfo{author}{van~der Geest, R.J.}, \bibinfo{year}{2018}.
\newblock \bibinfo{title}{Robust motion correction for myocardial {T1} and extracellular volume mapping by principle component analysis-based groupwise image registration}.
\newblock \bibinfo{journal}{Journal of Magnetic Resonance Imaging} \bibinfo{volume}{47}, \bibinfo{pages}{1397--1405}.
\bibitem[{Taylor et~al.(2016)Taylor, Salerno, Dharmakumar and Jerosch-Herold}]{taylor2016t1}
\bibinfo{author}{Taylor, A.J.}, \bibinfo{author}{Salerno, M.}, \bibinfo{author}{Dharmakumar, R.}, \bibinfo{author}{Jerosch-Herold, M.}, \bibinfo{year}{2016}.
\newblock \bibinfo{title}{{T1} mapping: basic techniques and clinical applications}.
\newblock \bibinfo{journal}{JACC: Cardiovascular Imaging} \bibinfo{volume}{9}, \bibinfo{pages}{67--81}.
\bibitem[{Teed and Deng(2020)}]{teed2020raft}
\bibinfo{author}{Teed, Z.}, \bibinfo{author}{Deng, J.}, \bibinfo{year}{2020}.
\newblock \bibinfo{title}{Raft: Recurrent all-pairs field transforms for optical flow}, in: \bibinfo{editor}{Vedaldi, A.}, \bibinfo{editor}{Bischof, H.}, \bibinfo{editor}{Brox, T.}, \bibinfo{editor}{Frahm, J.M.} (Eds.), \bibinfo{booktitle}{Proc. of Computer Vision--ECCV 2020: 16th European Conference, Glasgow, UK, August 23--28, 2020, Lecture Notes in Computer Science}, \bibinfo{organization}{Springer}. pp. \bibinfo{pages}{402--419}.
\bibitem[{Tilborghs et~al.(2019)Tilborghs, Dresselaers, Claus, Claessen, Bogaert, Maes and Suetens}]{tilborghs2019robust}
\bibinfo{author}{Tilborghs, S.}, \bibinfo{author}{Dresselaers, T.}, \bibinfo{author}{Claus, P.}, \bibinfo{author}{Claessen, G.}, \bibinfo{author}{Bogaert, J.}, \bibinfo{author}{Maes, F.}, \bibinfo{author}{Suetens, P.}, \bibinfo{year}{2019}.
\newblock \bibinfo{title}{Robust motion correction for cardiac {T1} and {ECV} mapping using a {T1} relaxation model approach}.
\newblock \bibinfo{journal}{Medical Image Analysis} \bibinfo{volume}{52}, \bibinfo{pages}{212--227}.
\bibitem[{Van~Zijl et~al.(1998)Van~Zijl, Eleff, Ulatowski, Oja, Uluǧ, Traystman and Kauppinen}]{van1998quantitative}
\bibinfo{author}{Van~Zijl, P.C.}, \bibinfo{author}{Eleff, S.M.}, \bibinfo{author}{Ulatowski, J.A.}, \bibinfo{author}{Oja, J.M.}, \bibinfo{author}{Uluǧ, A.M.}, \bibinfo{author}{Traystman, R.J.}, \bibinfo{author}{Kauppinen, R.A.}, \bibinfo{year}{1998}.
\newblock \bibinfo{title}{Quantitative assessment of blood flow, blood volume and blood oxygenation effects in functional magnetic resonance imaging}.
\newblock \bibinfo{journal}{Nature medicine} \bibinfo{volume}{4}, \bibinfo{pages}{159--167}.
\bibitem[{Wei et~al.(2024)Wei, Zhang, Zhang, Ding, Chen, Ong, Zhang and Xiang}]{wei2024memorization}
\bibinfo{author}{Wei, J.}, \bibinfo{author}{Zhang, Y.}, \bibinfo{author}{Zhang, L.Y.}, \bibinfo{author}{Ding, M.}, \bibinfo{author}{Chen, C.}, \bibinfo{author}{Ong, K.L.}, \bibinfo{author}{Zhang, J.}, \bibinfo{author}{Xiang, Y.}, \bibinfo{year}{2024}.
\newblock \bibinfo{title}{Memorization in deep learning: A survey}.
\newblock \bibinfo{journal}{arXiv preprint arXiv:2406.03880} .
\bibitem[{Weing{\"a}rtner et~al.(2015)Weing{\"a}rtner, Roujol, Ak{\c{c}}akaya, Basha and Nezafat}]{weingartner2015free}
\bibinfo{author}{Weing{\"a}rtner, S.}, \bibinfo{author}{Roujol, S.}, \bibinfo{author}{Ak{\c{c}}akaya, M.}, \bibinfo{author}{Basha, T.A.}, \bibinfo{author}{Nezafat, R.}, \bibinfo{year}{2015}.
\newblock \bibinfo{title}{Free-breathing multislice native myocardial {T1} mapping using the slice-interleaved {T1} ({STONE}) sequence}.
\newblock \bibinfo{journal}{Magnetic resonance in medicine} \bibinfo{volume}{74}, \bibinfo{pages}{115--124}.
\bibitem[{Xue et~al.(2012)Xue, Shah, Greiser, Guetter, Littmann, Jolly, Arai, Zuehlsdorff, Guehring and Kellman}]{xue2012motion}
\bibinfo{author}{Xue, H.}, \bibinfo{author}{Shah, S.}, \bibinfo{author}{Greiser, A.}, \bibinfo{author}{Guetter, C.}, \bibinfo{author}{Littmann, A.}, \bibinfo{author}{Jolly, M.P.}, \bibinfo{author}{Arai, A.E.}, \bibinfo{author}{Zuehlsdorff, S.}, \bibinfo{author}{Guehring, J.}, \bibinfo{author}{Kellman, P.}, \bibinfo{year}{2012}.
\newblock \bibinfo{title}{Motion correction for myocardial {T1} mapping using image registration with synthetic image estimation}.
\newblock \bibinfo{journal}{Magnetic resonance in medicine} \bibinfo{volume}{67}, \bibinfo{pages}{1644--1655}.
\bibitem[{Yang et~al.(2022)Yang, Zhao, Huang, Xia and Tao}]{yang2022disq}
\bibinfo{author}{Yang, C.}, \bibinfo{author}{Zhao, Y.}, \bibinfo{author}{Huang, L.}, \bibinfo{author}{Xia, L.}, \bibinfo{author}{Tao, Q.}, \bibinfo{year}{2022}.
\newblock \bibinfo{title}{{DisQ}: Disentangling quantitative {MRI} mapping of the heart}, in: \bibinfo{booktitle}{Proc. 2022 Int. Conf. Medical Image Computing and Computer-Aided Intervention (MICCAI 2022), Lecture Notes in Computer Science}, \bibinfo{organization}{Springer}. pp. \bibinfo{pages}{291--300}.
\bibitem[{Zhang et~al.(2018)Zhang, Le, Kabus, Su, Hausenloy, Cook, Chin and Tan}]{zhang2018cardiac}
\bibinfo{author}{Zhang, S.}, \bibinfo{author}{Le, T.T.}, \bibinfo{author}{Kabus, S.}, \bibinfo{author}{Su, B.}, \bibinfo{author}{Hausenloy, D.J.}, \bibinfo{author}{Cook, S.A.}, \bibinfo{author}{Chin, C.W.}, \bibinfo{author}{Tan, R.S.}, \bibinfo{year}{2018}.
\newblock \bibinfo{title}{Cardiac magnetic resonance {T1} and extracellular volume mapping with motion correction and co-registration based on fast elastic image registration}.
\newblock \bibinfo{journal}{Magnetic Resonance Materials in Physics, Biology and Medicine} \bibinfo{volume}{31}, \bibinfo{pages}{115--129}.
\bibitem[{Zhang et~al.(2024)Zhang, Zhao, Huang, Xia and Tao}]{zhang2024pca}
\bibinfo{author}{Zhang, Y.}, \bibinfo{author}{Zhao, Y.}, \bibinfo{author}{Huang, L.}, \bibinfo{author}{Xia, L.}, \bibinfo{author}{Tao, Q.}, \bibinfo{year}{2024}.
\newblock \bibinfo{title}{{PCA-R}elax: Deep-learning-based groupwise registration for motion correction of cardiac {T}$_1$ mapping}.
\newblock \bibinfo{journal}{arXiv preprint arXiv:2406.12456} .

\end{thebibliography}

\end{document}